\documentclass[twocolumn,prd,a4paper,final,superscriptaddress,longbibliography]{revtex4-1}
\usepackage{amsmath,amssymb}
\usepackage[polish,english]{babel}
\usepackage[latin1]{inputenc}
\usepackage{graphicx}
\usepackage{bbold}
\usepackage[T1]{fontenc}
\usepackage{epsfig}
\usepackage{epstopdf}
\usepackage{bbold}
\usepackage{comment}
\usepackage{bm}

\usepackage{natbib}

\graphicspath{{./fig/}}

\begin{document}
	
	\title{Gravitating compact $Q$-ball and $Q$-shell solutions in the $\mathbb{C}P^N$ nonlinear sigma model}
	
	\author{Pawe\l~Klimas}
	\email{pawel.klimas@ufsc.br}
	\affiliation{Universidade Federal de Santa Catarina, Trindade, 88040-900, Florian\'opolis, SC, Brazil}

	\author{Nobuyuki Sawado}
	\email{sawadoph@rs.tus.ac.jp}
	\affiliation{Department of Physics, Tokyo University of Science, Noda, Chiba 278-8510, Japan}

	\author{Shota Yanai}
	\email{phyana0513@gmail.com}
	\affiliation{Department of Physics, Tokyo University of Science, Noda, Chiba 278-8510, Japan}

	\vspace{.5 in}
	\small

	\date{\today}
	
	\begin{abstract}
	We study compact gravitating $Q$-ball, $Q$-shell solutions in a sigma model with the target space $\mathbb{C}P^N$. 
	Models with odd integer $N$ and suitable potential can be parameterized by $N$-th complex scalar fields and they 
	support compact solutions.  
	A coupling with gravity allows for harboring of the Schwarzschild black holes for the $Q$-shell 
	solutions. The energy of the solutions behaves as $E\sim |Q|^{5/6}$, where $Q$ stands for the $U(1)$ Noether charge,  
	for both the gravitating and the black hole solutions.
	Notable difference from the solutions of the flat space is that 
	upper bound of $|Q|$ appears when the coupling with gravity is stronger.
	The maximal value of $|Q|$ quickly reduces for larger coupling constant.
	It may give us a useful hint of how a star forms its shape with a certain finite number of particles.  
	\end{abstract}
	
	\pacs{}

	\maketitle 
	\section{Introduction}
	\label{Intro}
		
	$Q$-balls are stationary soliton solutions of certain complex scalar field theories with self-interactions
	\citep{Friedberg:1976me,Coleman:1985ki,Leese:1991hr,Loiko:2018mhb}. 
	The U(1) invariance of the scalar field 
	leads to the conserved charge $Q$ which is identified with the electric charge of the constituents for theories coupled to 		
	electromagnetic field. 
	$Q$-balls have attracted much attention in the studies of 
	evolution of the early Universe \citep{Friedberg:1986tq,Lee:1986ts}.
	In supersymmetric extensions of the standard model, $Q$-balls appear as
	the scalar superpartners of baryons or leptons and they form coherent states with 
	baryon or lepton number. They may survive as a major ingredient of dark matter
	\citep{Kusenko:1997zq,Kusenko:1997si,Kusenko:1997vp}.  
	
	In order to get the standard $Q$-balls  one usually considers solutions with 
	constant angular velocity $\omega$ in the internal space of fields and with spherical symmetry 
	in position space. An absolute value of the Noether charge $|Q|$ has no upper bound while 
	the angular velocity always has some limitations. 
	The solutions exist when the charge $Q$ is greater than the lower critical value $Q_{\rm C}$.
	When concerning the absolute stability then another critical point emerges. Namely, a $Q$-ball is stable against decay into 		
	smaller $Q$-balls (single consituents) if its energy, above the critical value $Q_S$, is less then $mQ$, where $m$ is the mass 	
	of a single constituent.

	Some are other types of $Q$-balls associated with local symmetries have been extensively studied 
	in \citep{Lee:1988ag,Anagnostopoulos:2001dh,Kusenko:1997vi}.
	Such $Q$-balls are unstable if they possess  sufficiently large charge. The instability is caused by repulsion that 				
	originates in the gauge force. It means that for stable solutions the charge $|Q|$ has upper bound.  
	
	Compactons are field configurations that exist on finite-size supports. Outside this support the field is identically zero. 
	In the 	case of spherical configurations the support is delimited by outer radius $r_{\rm out}$. 
	For instance, the signum-Gordon model i.e. the scalar field model with standard kinetic terms and $V-$shaped 
	potential gives rise to such solutions \citep{Arodz:2008jk,Arodz:2008nm}. Interestingly, 
	when the scalar field is coupled with electromagnetism then the inner radius emerges, 
	i.e., the scalar field vanishes also in the central region $r<r_{\rm in}$. 
	Thus, the matter field exists in the region $r_{\rm in}\leqq r \leqq r_{\rm out}$. Such configurations of fields are called $Q$-shells. 
	Such shell solutions have no restrictions on upper bound for $|Q|$.
	The authors claim that the energy of compact $Q-$balls scales as $\sim Q^{5/6}$
	and of $Q$-shells for large $Q$ as $\sim Q^{7/6}$. 
	It clearly indicates that the $Q$-balls are stable against the decay while the $Q$-shells may be 
	unstable. 
	
	Many results concerning compact boson stars and shells are presented in \citep{Kleihaus:2010ep,Kumar:2014kna,Kumar:2015sia,Kumar:2016sxx}.
	For the boson shell configurations, one possibility is the case that the gravitating boson 
	shells surround a flat Minkowski-like interior region $r<r_{\rm in}$ while the exterior
	region $r>r_{\rm out}$ is the exterior of an Reissner-Nordstr\"om solution. 
	Another and even more interesting possibility is the existence of the charged black hole 
	in the interior region. The gravitating boson shells can harbor a black hole. 
	Since the black hole is surrounded by a shell of scalar fields, such
	fields outside of the event horizon may be interpreted as a scalar hair. 	
	Such possibility has been considered as contradiction of the no-hair conjecture.  
	The higher dimensional generalizations have been considered in \citep{Hartmann:2012da,Hartmann:2013kna}.
		
	Recently, a new model supporting both $Q$-balls and $Q$-shells has been proposed by one of us \citep{Klimas:2017eft}. 
	The  model is defined in 3+1 dimensions and it has the Lagrangian density
	\begin{align}
	{\cal L}=-\frac{M^2}{2}\mathrm{Tr}(X^{-1}\partial_\mu X)^2-\mu^2V(X),\label{lag0}
	\end{align}
	where the coupling constants $M$, $\mu$ have dimensions of $(\mathrm{length})^{-1}$ and $(\mathrm{length})^{-2}$, respectively.  
	The potential $V$ is chosen in the way that the model supports compact solutions. For models with standard kinetic terms 
	(it is the case of model \eqref{lag0}) it is sharp at its minima  and so it is called V-shaped potential. 
	The model \eqref{lag0} is a direct extension of the $\mathbb{C}P^1$ nonlinear sigma model on a model with target space $\mathbb{C}P^N$. 
	The field $X$ is called the principal variable and it successfully parameterizes the coset space
	 $SU(N+1)/U(N)\sim \mathbb{C}P^N$.
	There exist compact solutions for all odd numbers $N$,i.e., $N:=2n+1,n=0,1,2,\cdots$. 
	The salient feature of the model from other models containing only $Q$-balls, is that it
	supports both $Q$-ball and $Q$-shell solutions and the existence of $Q$-shells does not
	require the electromagnetic field.  
	
	For $n=0,1$ the solutions form $Q$-ball while for $n\geqq 2$ they always are $Q$-shell like.
	Again, there is no upper bound for the charge $|Q|$ and also there is no limitation from above for $\omega$. 
	In the complex signum-Gordon model with local symmetry  $Q$-balls exist due to 
	presence of the gauge field whereas in the case of the $\mathbb{C}P^N$ model 
	they appear as the result of self-interactions between scalar fields. 
	
	In this paper, we consider model containing complex scalar fields coupled to gravity and obtain the compact $Q$-ball and 
	$Q$-shell solutions. 
	The resulting self-gravitating regular solutions can be identified with boson stars \citep{Tamaki:2010zz}. 
	We also study  the systems composed of compact $Q$-shell solutions and the Schwarzschild-like black holes located in the 	
	center of the shell. The space-time inside $r<r_{\rm in}$ and outside the shell $r>r_{\rm out}$ is the Schwarzschild space-time. 
	The systems with these property might be considered as possible examples of violation of the no-hair conjecture. 
	A notable difference between these solutions and solutions  in flat space-time is that the first ones possess 
	upper bound for $|Q|$ and the value of this bound drastically decreases with increasing of gravitational coupling constant. 
	The maximal charge $|Q|$ is attained for large $n$ for sufficiently small gravitational constant which
	explain, to some extent, why the boson stars have definite size and mass. 
	In this paper we only present results for the Schwarzschild space-time, however,  
	the analysis can be extended on the case of electrically charged solutions with Reissner-Nordstr\"om metric.   

	The paper is organized as follows. In Section II we describe the model,
	coupled to gravity and discuss its parametrization.  In Section III we present 
	numerical results. Section IV is devoted to detailed analysis of  stability of the $Q$-ball, $Q$-shell solutions.
	Conclusions and remarks are given in the last Section.

	\section{The model}
	\label{model}
	
	\subsection{The action, the equations of motion}
	We consider the action of a self-gravitating complex variable $X$ coupled
	to Einstein gravity
	\begin{align}
	S=\int\sqrt{-g}d^4x\Bigl[
	\frac{R}{16\pi G}-\frac{M^2}{2}\mathrm{Tr}(X^{-1}\partial_\mu X)^2-\mu^2V(X)
	\Bigr]
	\label{action}
	\end{align}
	where $G$ is Newton's gravitational constant. 
	The $\mathbb{C}P^N$ space has a nice parametrization in terms of the principal 
	variable $X$ \citep{Eichenherr:1979hz} (see
	\citep{Ferreira:2010jb,Klimas:2017eft} for explicit construction of the solutions), defined as
	\begin{align}
	X(g)=g\sigma(g)^{-1},~~~~g\in SU(N+1)\,.
	\end{align}
	It parametrizes the coset space $\mathbb{C}P^N=SU(N+1)/SU(N)\otimes U(1)$ with the subgroup
	$SU(N)\otimes U(1)$ being invariant under the involutive automorphism $(\sigma^2=1)$.
	It follows that $X(gk)=X(g)$ for $\sigma(k)=k,~~k\in SU(N)\otimes U(1)$.
	The $\mathbb{C}P^N$ model possesses some symmetries.  
	It is easy to see that the kinetic term of the Lagrangian has the symmetry $X\to AXB^\dagger,~A,B\in SU(N+1)$. Thus the potential 	
	is chosen so as to break the symmetry down to the diagonal one, i.e., $X\to AXA^\dagger$. 
	The existence of compact solutions require special class of potentials. An 
	example of such a potential which we shall adopt in this paper has the form
	\begin{align}
	V(X)=\frac{1}{2}[\mathrm{Tr}(I-X)]^{1/2}\,.
	\end{align}
	As we shall see later, the behavior of fields at the outer border of compacton (and also at spatial infinity) implies $X\to I$.

	We assume the $(N+1)$-dimensional representation in which the $SU(N+1)$ valued group element $g$
	is parametrized by the set of complex fields $u_i$:
	\begin{align}
	g\equiv 
	\frac{1}{\vartheta^2}
	\left(\begin{array}{cc}
	\Delta & iu \\
	iu^\dagger & 1  
	\end{array}\right),~~~~
	\Delta_{ij}\equiv \vartheta \delta_{ij}-\frac{u_iu_j^*}{1+\vartheta}
	\label{parasu3}
	\end{align}
	which fixes the symmetry to $SU(N)\otimes U(1)$. 
	The principal variable takes the form
	\begin{align}
	X(g)=
	\left(\begin{array}{cc}
	I_{N\times N} & 0 \\
	0 & -1 
	\end{array}\right)+
	\frac{2}{\vartheta^2}
	\left(\begin{array}{cc}
	-u\otimes u^\dagger & iu \\
	iu^\dagger & 1  
	\end{array}\right)
	\label{principalu}
	\end{align}
	where $\vartheta:=\sqrt{1+u^\dagger\cdot u}$. 		
	Thus the $\mathbb{C}P^N$ Lagrangian of the model (\ref{action}) 
	takes the form
	\begin{align}
	{\cal L}_{\mathbb{C}P^N}=-M^2g^{\mu\nu}\tau_{\nu\mu}-\mu^2 V
	\label{lagrangianu}
	\end{align}
	where
	\begin{align}
	\tau_{\nu\mu}=-\frac{4}{\vartheta^4}\partial_\mu u^\dagger\cdot \Delta^2\cdot \partial_\nu u,~~
	\Delta_{ij}^2:=\vartheta^2\delta_{ij}-u_iu_j^*\,.
	\end{align}
	
	The variation of the action with respect to the metric leads to Einstein's equations
	\begin{align}
	G_{\mu\nu}=8\pi GT_{\mu\nu},\qquad{\rm where}\qquad G_{\mu\nu}\equiv R_{\mu\nu}-\frac{1}{2}g_{\mu\nu}R
	\label{einstein_formal}
	\end{align}
	and where the stress-energy tensor is of the form
	\begin{align}
	T_{\mu\nu}&=2\frac{\partial {\cal L}_{\mathbb{C}P^N}}{\partial g^{\mu\nu}}-g_{\mu\nu}{\cal L}_{\mathbb{C}P^N} \nonumber \\
	&=-2M^2\tau_{\nu\mu}+M^2g_{\mu\nu}g^{\alpha\beta}\tau_{\beta\alpha}+g_{\mu\nu}\mu^2V\,.
	\label{stress_formal}
	\end{align}
	Next, varying the action with respect to fields $u_i$ and $u_i^*$ one obtains the $\mathbb{C}P^N$ field equations
	\begin{align}
	&\frac{1}{\sqrt{-g}}\partial_\mu(\sqrt{-g}\partial^\mu u_i)+\frac{2}{\vartheta^2}(\partial_\mu u^\dagger\cdot u)\partial^\mu u_i
	\nonumber \\
	&\hspace{1cm}+\frac{\mu^2}{4M^2}\vartheta^2\sum_{k=1}^N\Bigl[
	(\delta_{ik}+u_iu_k^*)\frac{\partial V}{\partial u_k^*}
	\Bigr]=0\,.
	\label{cpneq_formal}
	\end{align}
	The obtained system of coupled equations \eqref{einstein_formal} and \eqref{cpneq_formal} is quite complex so we shall 
	adopt numerical techniques to solve it. 
	
	It is convenient to introduce the the dimensionless variables
	\begin{align}
	x_\mu \to \frac{\mu}{M} x_\mu\,.
	\end{align}
	For the solutions with spherical symmetry, we employ the standard Schwarzschild-like coordinates such that the line element reads
	\begin{align}
	ds^2&=g_{\mu\nu}dx^\mu dx^\nu \nonumber \\
	&=A^2(r)C(r)dt^2-\frac{1}{C(r)}dr^2-r^2(d\theta^2+\sin^2\theta d\varphi^2)
	\label{metric}
	\end{align}
	where 
	\begin{align}
	C(r):=1-\frac{2m(r)}{r}\,.
	\end{align}

	We also restrict $N$ to be odd, i.e., $N:=2n+1$.  The ansatz
	\begin{align}
	u_m(t,r,\theta,\varphi)=\sqrt{\frac{4\pi}{2n+1}}f(r)Y_{nm}(\theta,\varphi)e^{i\omega t}
	\label{ansatz}
	\end{align}
	allows for reduction of partial equations to the system of radial ordinary equations, 
	where $Y_{nm}, -n\leq m \leq n$ are the standard spherical harmonics and $f(r)$ is the profile function. 
	Each $2n+1$ field $u=(u_m)=(u_{-n},u_{-n+1},\cdots,u_{n-1},u_n)$ is associated with one of $2n+1$
	spherical harmonics for given $n$. 
	The relation 
	$\sum_{m=-n}^n Y_{nm}^*(\theta,\varphi)Y_{nm}(\theta,\varphi)=\frac{2n+1}{4\pi}$
	is very useful for obtaining an explicit form of many inner products. 
	Substituting (\ref{metric}) and (\ref{ansatz}) into the Einstein field equations (\ref{einstein_formal}),
	we get their components  
	\begin{align}
	&(t~t):~~
	\frac{[r(1-C)]'}{r^2}= \nonumber \\
	&~~~~\alpha\biggl(
	\frac{4\omega^2f^2}{A^2C(1+f^2)^2}+\frac{4Cf'^2}{(1+f^2)^2}+\frac{4n(n+1)f^2}{r^2(1+f^2)}+\frac{f}{\sqrt{1+f^2}} 
	\biggr)\,,
	\label{einsteint}
	\\
	&(r~r):~~
	\frac{-A[r(1-C)]'+2rA'C}{r^2A}= \nonumber \\
	&~~~~\alpha\biggl(
	\frac{4\omega^2f^2}{A^2C(1+f^2)^2}+\frac{4Cf'^2}{(1+f^2)^2}-\frac{4n(n+1)f^2}{r^2(1+f^2)}-\frac{f}{\sqrt{1+f^2}} 
	\biggr)\,,
	\label{einsteinr}
	\\
	&(\theta~\theta):~~
	\frac{3rA'C'+2C(rA')'-A[r(1-C)]''}{2rA}= \nonumber \\
	&~~~~\alpha\biggl(
	\frac{4\omega^2f^2}{A^2C(1+f^2)^2}-\frac{4Cf'^2}{(1+f^2)^2}-\frac{f}{\sqrt{1+f^2}} 
	\biggr)\,.
	\label{einsteinth}
	\end{align}
	Then from (\ref{einsteint}) and (\ref{einsteinr}) we obtain equations for the metric tensor functions $A$ and $C$ 
	\begin{align}
	&A'=\alpha r A\biggl(
	\frac{4\omega^2f^2}{A^2C^2(1+f^2)^2}+\frac{4f'^2}{(1+f^2)^2}
	\biggr)\,,
	\label{einsteinA}
	\\
	&C'=\frac{1-C}{r} \nonumber \\
	&-\alpha r\biggl(
	\frac{4\omega^2f^2}{A^2C(1+f^2)^2}+\frac{4Cf'^2}{(1+f^2)^2}+\frac{4n(n+1)f^2}{r^2(1+f^2)}+\frac{f}{\sqrt{1+f^2}} 
	\biggr)
	\label{einsteinC}
	\end{align}
	where $\alpha$ is a dimensionless coupling constant defined as 
	$$\alpha:=8\pi G\mu^2.$$
	We shall solve numerically such obtained equations in dependence on
	the coupling constant $\alpha$ and the frequency $\omega$. 
	Variation of the parameter $\alpha$ is equivalent to variation of the model parameter $\mu$ (or of $M$ in fully dimensional case). 
	In this context, the parameter $\mu$ is a kind ``susceptibility'' of gravity to matter. 

\begin{figure*}[t]
  \begin{center}
    \includegraphics[width=80mm]{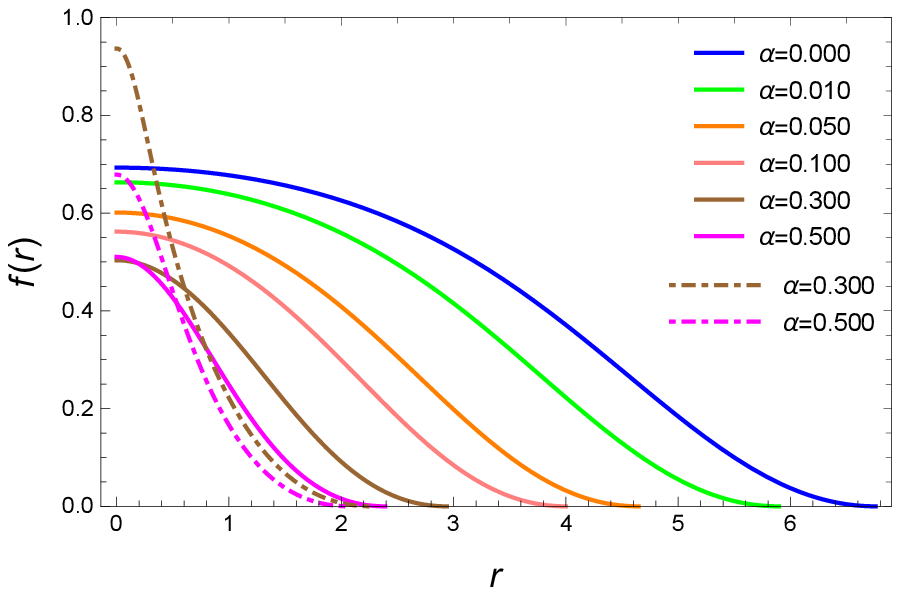}~~
    \includegraphics[width=80mm]{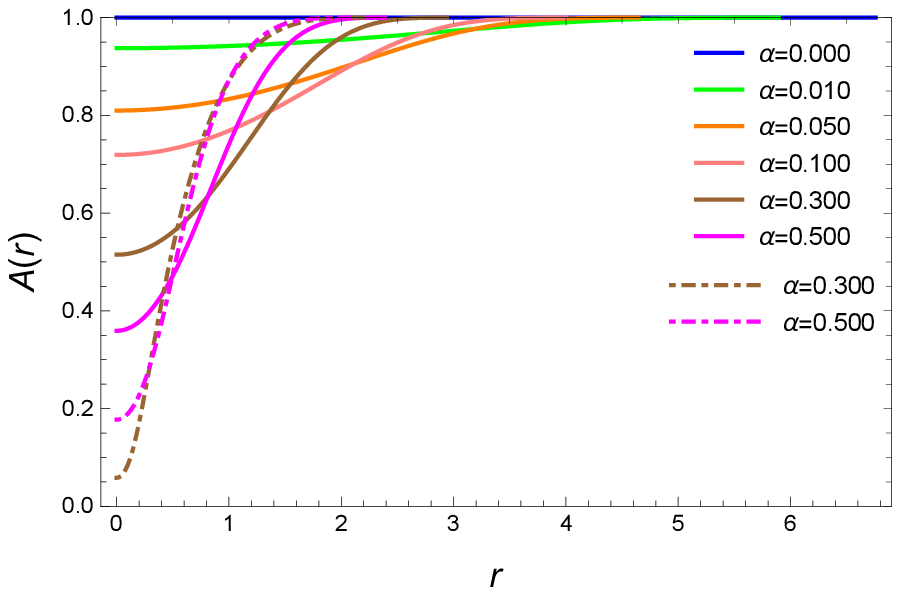}\\
    \vspace{5mm}

    \includegraphics[width=80mm]{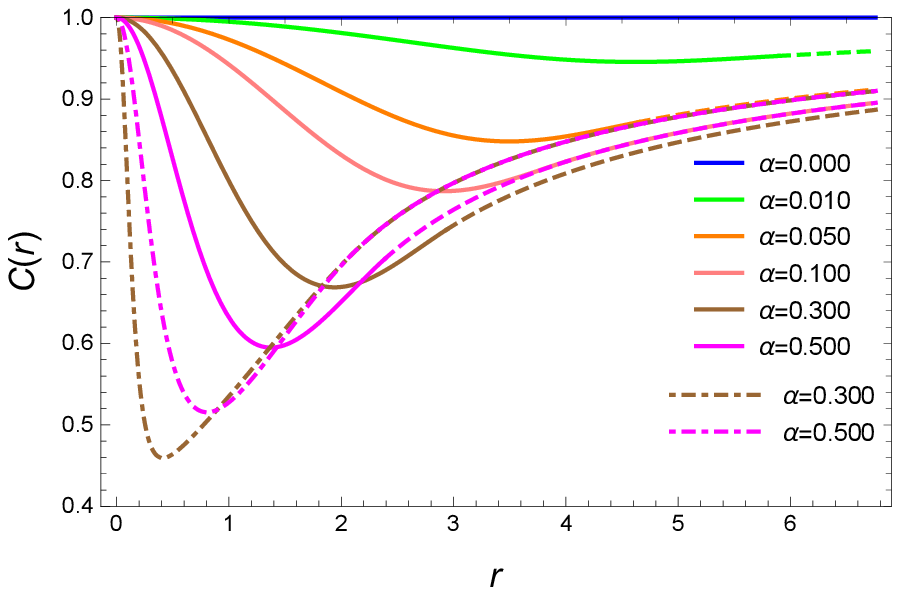}~~
    \includegraphics[width=80mm]{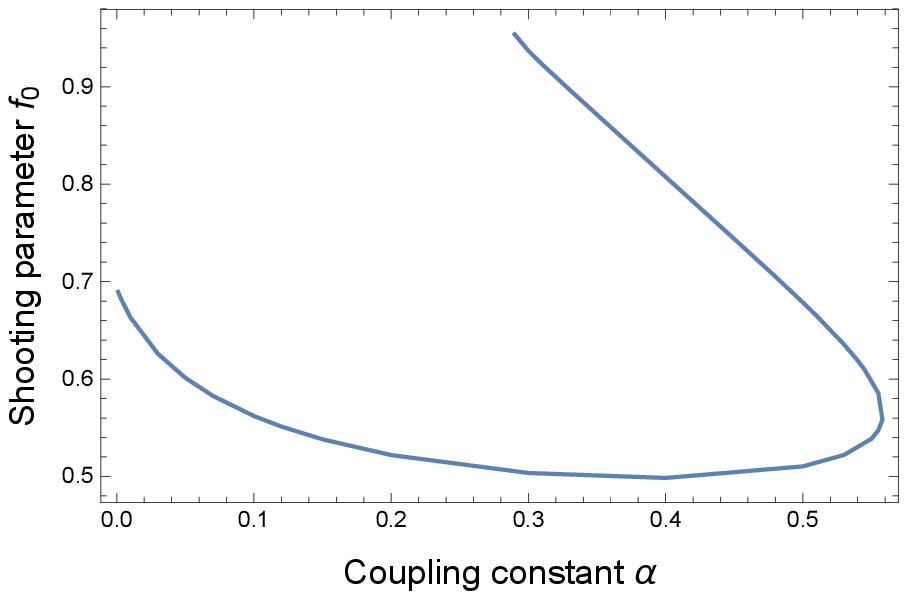}
    \caption{\label{CP1}The gravitating $Q$-ball solution for the $\mathbb{C}P^{1}$ case. The matter profile function (top left). The 
    straight lines represent the stable branch whereas the dotted lines stand for the unstable branch.  The metric function $A(r)$
    (the top right figure) and the metric function $C(r)$ (the bottom left figure). 
	Distinct curves correspond with different values of coupling constant $\alpha$.
    The bottom right figure shows a relation between shooting parameter $f_0$ and coupling constant $\alpha$.}
  \end{center}
	\end{figure*}

\begin{figure*}[t]
  \begin{center}
\includegraphics[width=80mm]{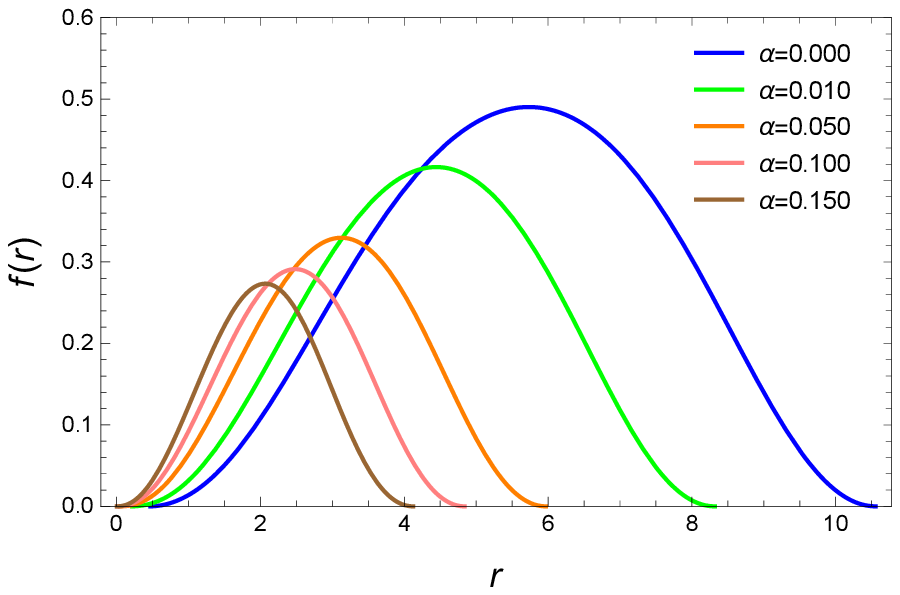}~~
\includegraphics[width=80mm]{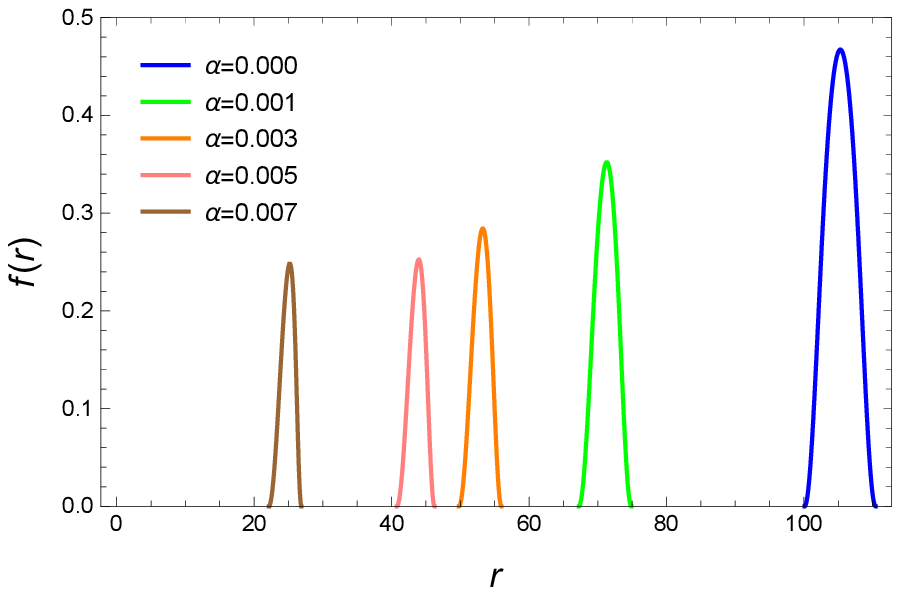} \\
\vspace{5mm}

\includegraphics[width=80mm]{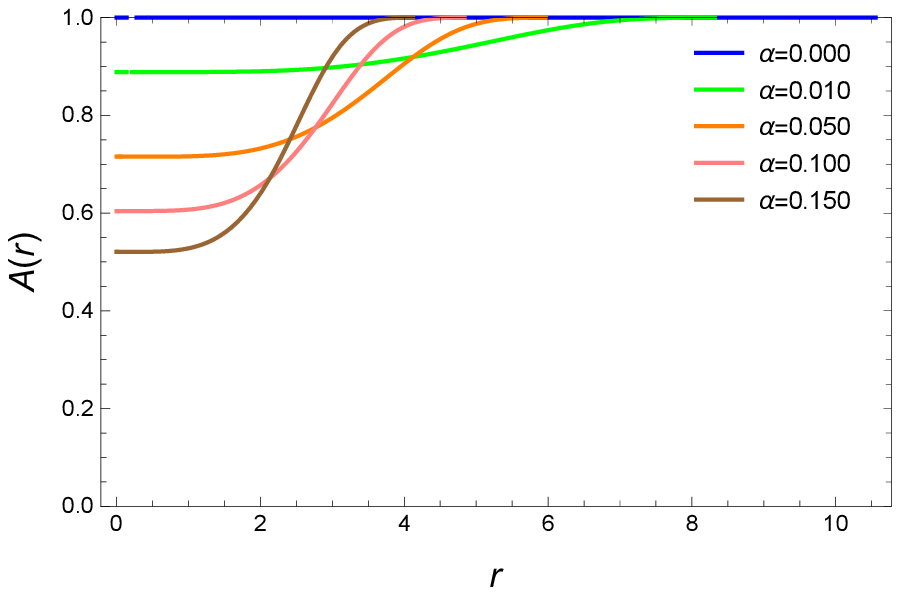}~~
\includegraphics[width=80mm]{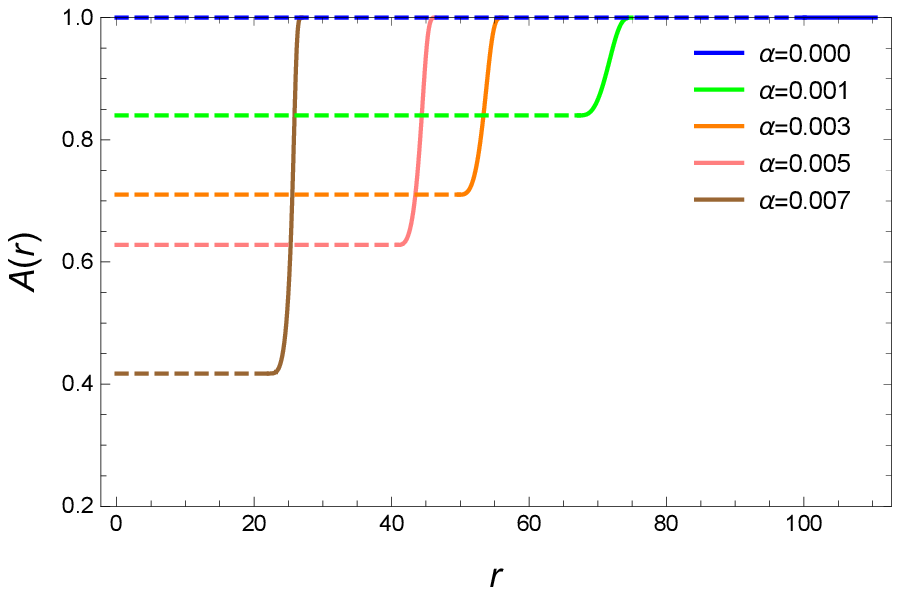} \\
\vspace{5mm}

\includegraphics[width=80mm]{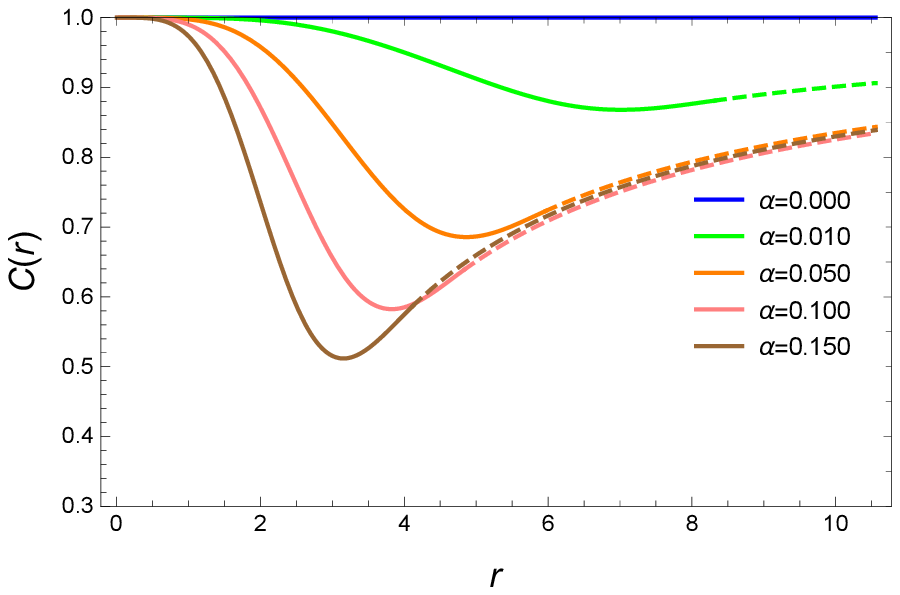}~~
\includegraphics[width=80mm]{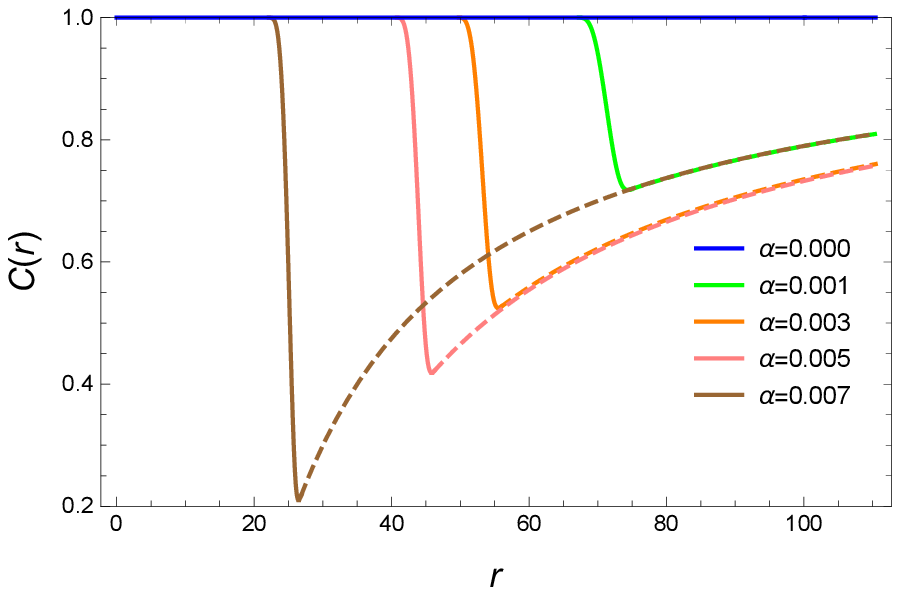}

    \caption{\label{CPgra}The self-gravitating  $Q$-shell solutions for the case $\mathbb{C}P^5$ (the left figures) and for $\mathbb{C}P^{101}$ (the right figures):
the matter profile functions (the top figures), the metric functions $A(r)$ (the central figures) and the $C(r)$ (the bottom figures). The dotted line indicates solutions of the vacuum Einstein equations.}
  \end{center}
\end{figure*}

	The $\mathbb{C}P^N$ equations (\ref{cpneq_formal}) read  
	\begin{align}
	&Cf''+\frac{2Cf'}{r}+C'f'+\frac{A'Cf'}{A}-\frac{n(n+1)f}{r^2} \nonumber \\
	&+\frac{\omega^2f(1-f^2)}{A^2C(1+f^2)}-\frac{2Cff'^2}{1+f^2}
	-\frac{1}{8}\mathrm{sgn}(f)\sqrt{1+f^2}=0\,.
	\label{matter}
	\end{align}
	Here the prime $ ' $ means differentiation with respect to $r$.

	The dimensionless Lagrangian 
	\begin{align}
	\tilde{{\cal L}}_{\mathbb{C}P^N}=-g^{\mu\nu}\tau_{\nu\mu}-V\,,
	\label{lagdimless}
	\end{align}
	allows us to obtain the dimensionless Hamiltonian of the $\mathbb{C}P^N$ model, which reads
	\begin{align}
	{\cal H}_{\mathbb{C}P^N}	
	&=-g^{00}\tau_{00}+g^{ii}\tau_{ii}+ V \nonumber \\
	&=\frac{4}{(1+f^2)^2}\biggl(Cf'^2
	+\frac{\omega^2f^2}{A^2C}+\frac{n(n+1)(1+f^2)f^2}{r^2}
	\biggr) \nonumber \\
 	&~~~~+\frac{f}{\sqrt{1+f^2}}.
	\end{align}
	 The symmetry $SU(N+1)$ is reduced to $SU(N)\otimes U(1)$ for parametrization (\ref{parasu3}). It is the symmetry of
	the Lagrangian (\ref{lagdimless}). 
	It contains subgroup $U(1)^N$ given by the transformation
	\begin{align}
	u_i\to e^{i\beta_i}u_i,~~i=1,\cdots,N
	\label{u1}
	\end{align}
	where $\beta_i$ are some global parameters. The Noether charge associated with this global symmetry plays important role for stability of solutions. 
	The Noether currents corresponding with the symmetry transformation (\ref{u1}) read
	\begin{align}
	J_\mu^{(i)}=-\frac{4i}{\vartheta^4}\sum_{j=1}^N[u_i^*\Delta_{ij}^2\partial_\mu u_j-\partial_\mu u_j^*\Delta_{ji}^2u_i]\,.
	\end{align}
	Using the ansatz ($\ref{ansatz}$) we find following form of the Noether currents
	\begin{align}
	&J^{(m)}_0=8\omega\frac{(n-m)!}{(n+m)!}\frac{f^2}{(1+f^2)^2}(P_n^m(\cos\theta))^2\,,
	\label{currentt}\\
	&J^{(m)}_\varphi=8m\frac{(n-m)!}{(n+m)!}\frac{f^2}{1+f^2}(P_n^m(\cos\theta))^2
	\label{currentp}
	\end{align}
	and $J_r^{(m)}=J_\theta^{(m)}=0$, for $m=-n,-n+1,\cdots,n-1,n$. 
	The conservation of currents is almost straightforward because non-vanishing components (\ref{currentt}) and (\ref{currentp}) depend only on 
	$r$ and $\theta$.  Thus 
	\begin{align}
	&\frac{1}{\sqrt{-g}}\partial_\mu(\sqrt{-g}g^{\mu\nu} J_\nu^{(m)}) \nonumber \\
	&~~~~=\frac{1}{A^2C}\partial_0J_0^{(m)}+\frac{1}{r^2\sin^2\theta}\partial_\varphi J_\varphi^{(m)}=0\,.
	\label{current_cons}
	\end{align}
	The integral of (\ref{current_cons}) over $[t',t'']\times {\mathbb R^3}$ gives 
	\begin{align}
	0=\int_{t'}^{t''} dt\int_{\mathbb R^3} d^3x \sqrt{-g}\biggl(
	\frac{1}{A^2C}\partial_0J_0^{(m)}+\frac{1}{r^2\sin^2\theta}\partial_\varphi J_\varphi^{(m)}
	\biggr)\,.
	\end{align}

	When $J_\varphi^{(m)}$ decreases sufficiently quickly at the spatial boundary (and so it does not contributes to the integral) the Noether charge
	\begin{align}
	Q^{(m)}&=\frac{1}{2}\int_{\mathbb R^3} d^3 x\sqrt{-g}\frac{1}{A^2C}J_0^{(m)}(x) \nonumber \\
	&=\frac{16\pi\omega}{2n+1}\int^\infty_0r^2dr\frac{f^2}{AC(1+f^2)^2}\,
	\end{align}
	is conserved.
	The spatial components of the Noether currents do not contribute to the charges, however,
	they can be used to introduce some auxiliary integrals
	\begin{align}
	q^{(m)}&:=\frac{3}{2}\int d^3x\sqrt{-g}\frac{J^{(m)}_\varphi (x)}{r^2} \nonumber \\
	&=\frac{48\pi m}{2n+1}\int^\infty_0dr\frac{Af^2}{1+f^2}\,.
	\end{align}
	With the help of $Q^{m}$ and $q^{m}$ the total energy $E$ can be expressed in the form
	\begin{align}
	E=4\pi\int r^2 drA\biggl(\frac{4Cf'^2}{(1+f^2)^2}+\frac{f}{\sqrt{1+f^2}}\biggr) \nonumber \\
	+\sum_{m=-n}^{m=n}(\omega Q^{(m)}+m q^{(m)})\,.
	\end{align}

\begin{figure*}[t]
  \begin{center}
\includegraphics[width=80mm]{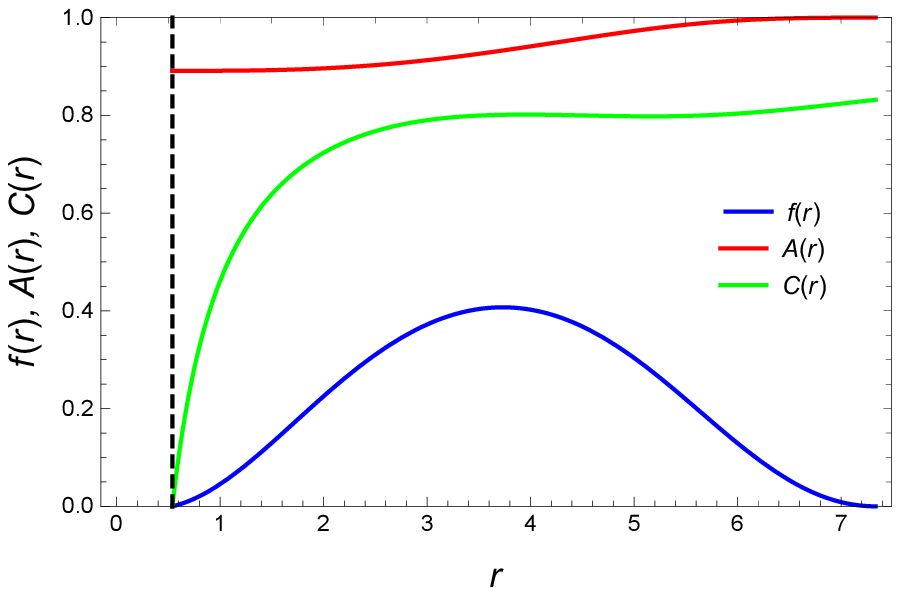}~~
\includegraphics[width=80mm]{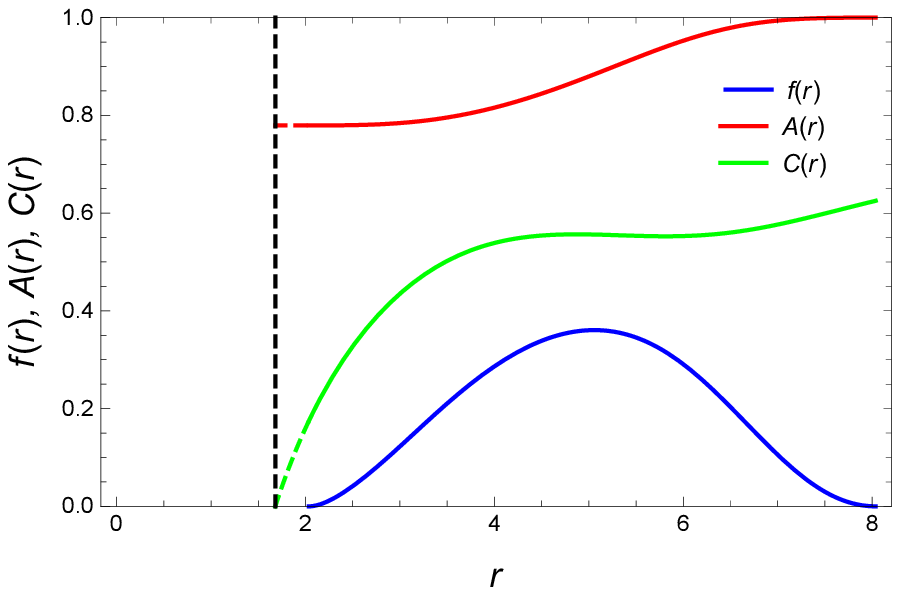} \\
\vspace{5mm}

\includegraphics[width=80mm]{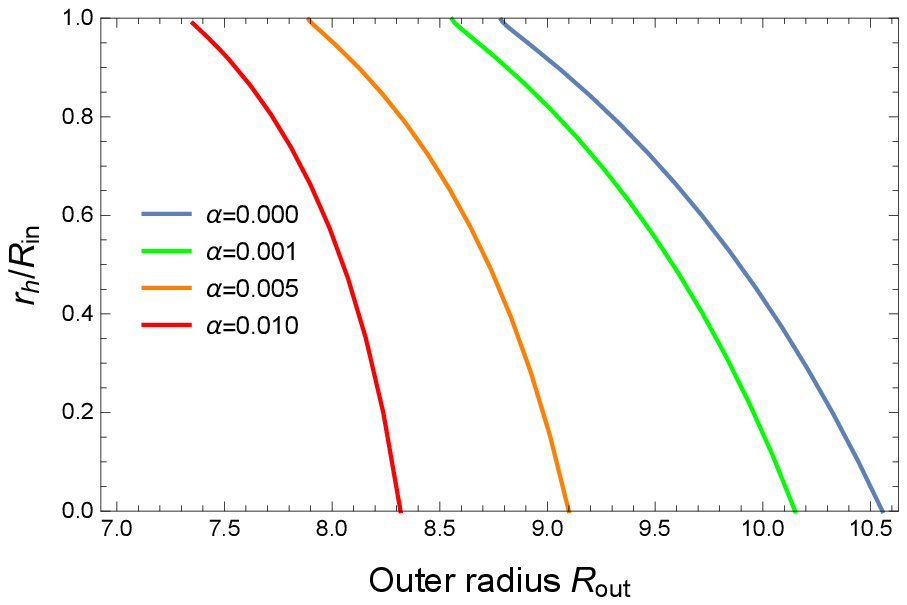}~~
\includegraphics[width=80mm]{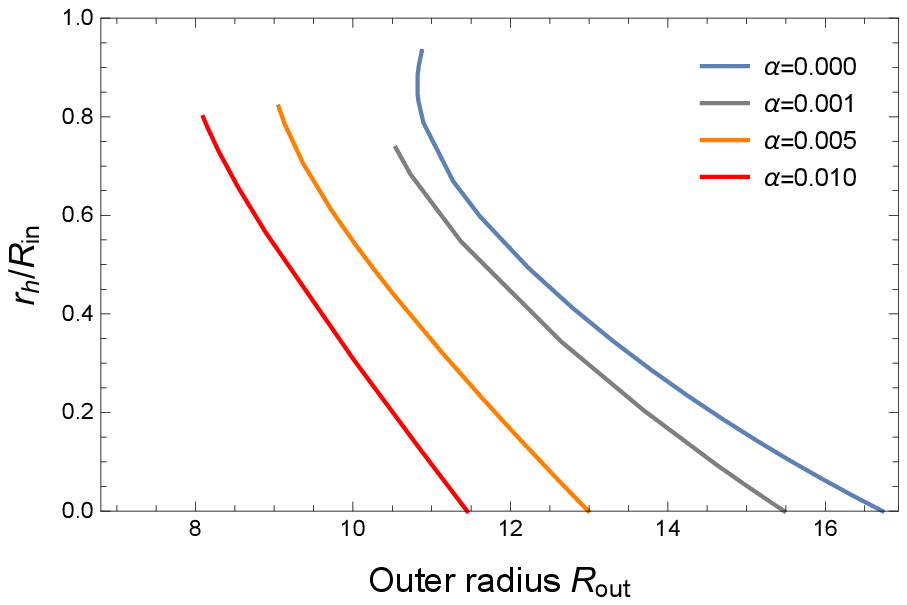} \\
\vspace{5mm}

\includegraphics[width=80mm]{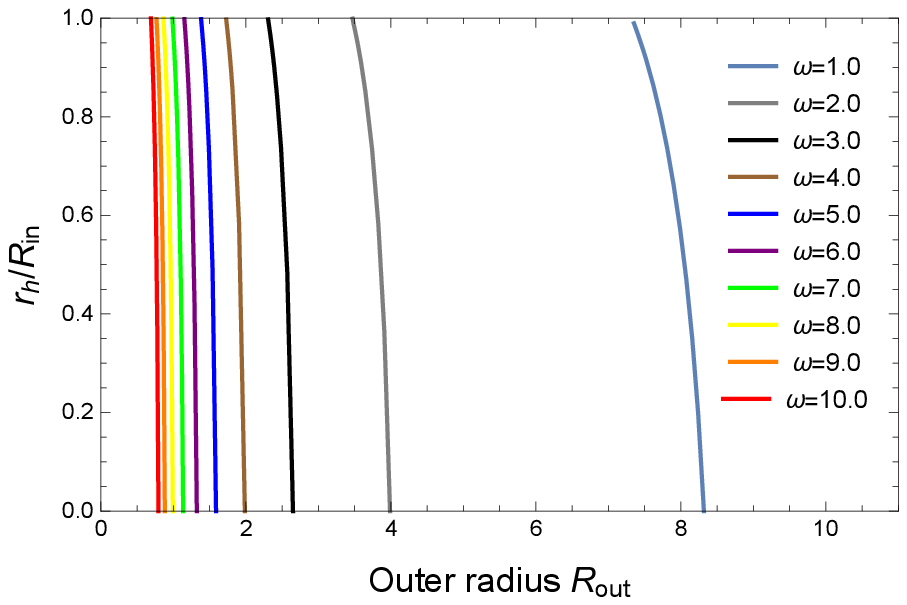}~~
\includegraphics[width=80mm]{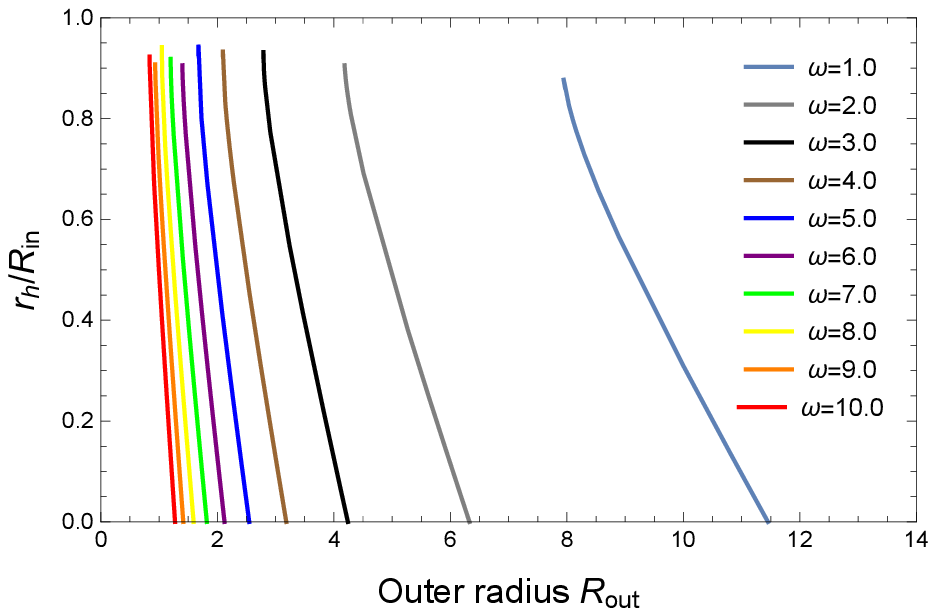}

    \caption{\label{CPBH}The harbor solutions for the $\mathbb{C}P^5$ (the left column figures) and the $\mathbb{C}P^{11}$ (the right column figures): 
the radius of the event horizon is chosen as $r_{\rm h}=0.5385$ ($\mathbb{C}P^5$) and $r_{\rm h}=1.680$ ($\mathbb{C}P^{11}$).
The central two figures 
plot the ratio of the event horizon/inner radius $r_{\rm h}/R_{\rm in}$  as a function of outer radius $R_{\rm out}$
for several of the coupling constant $\alpha$ and fixed $\omega=1.0$. 
The bottom two figures show the ratio of the radii for several values of $\omega$ and fixed $\alpha=0.01$. }
  \end{center}
\end{figure*}

	\subsection{The behavior of solutions at the boundary}
	
	In this section we shall discuss behavior of solutions at the boundary, which means that we mainly look at the origin $r=0$ and the border(s) of the compacton. 
	First we consider expansion at the origin and so the solution is represented by series
	\begin{align}
	&f(r)=\sum_{k=0}^\infty f_kr^k,~~~~
	A(r)=\sum_{k=0}^\infty A_kr^k, \nonumber \\
	&m(r)=\sum_{k=0}^\infty m_kr^k .
	\end{align}
	After substituting these expressions into equations (\ref{einsteinA}), (\ref{einsteinC}), (\ref{matter}), one requires vanishing of equations in all orders of expansion. It allows us to determinate the coefficients of expansion. The form of coefficients depends on the value of parameter $n$.
	For $n=0$ it reads
	\begin{align}
	&f(r)=f_0+\frac{1}{48}\biggl(\sqrt{1+f_0^2}-\frac{8f_0(1-f_0^2)\omega^2}{A_0^2(1+f_0^2)}\biggr)r^2+O(r^4)\,, \nonumber \\
	&A(r)=A_0+\frac{2\alpha f_0^2\omega^2}{A_0(1+f_0^2)^2}r^2+O(r^4)\,,\label{expansion0} \\
	&m(r)=\frac{1}{6}\biggl(\frac{\alpha f_0}{(1+f_0)^{1/2}}+\frac{4f_0^2\omega^2}{A_0^2(1+f_0^2)^2}\biggr)r^3+O(r^4) \nonumber
	\end{align}
	where $f_0$, $ A_0$ are free parameters.
	For $n=1$ we obtain
	\begin{align}
	&f(r)=f_1r+\frac{1}{32}r^2+\frac{1}{10}\biggl(2f_1^3(1+7\alpha)-\frac{f_1\omega^2}{A_0^2}\biggr)r^3+O(r^4)\,, \nonumber \\
	&A(r)=A_0+\alpha A_0f_1^2r^2+\frac{1}{12}\alpha A_0f_1r^3+O(r^4)\,,\label{expansion1} \\
	&m(r)=2\alpha f_1^2r^3+O(r^4) \nonumber
	\end{align}
	with free parameters $f_1$ and $A_0$.

\begin{figure*}[t]
  \begin{center}
\includegraphics[width=70mm]{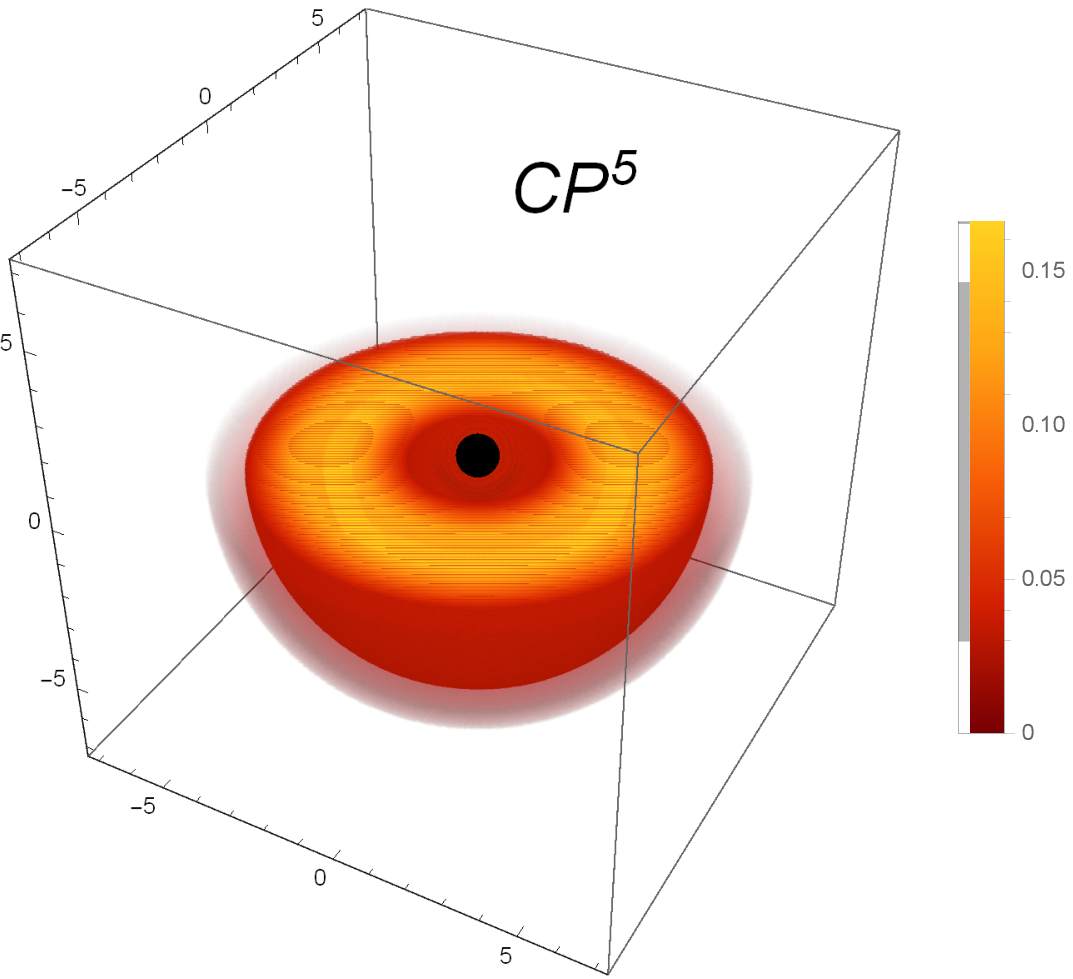}~~~~
\includegraphics[width=70mm]{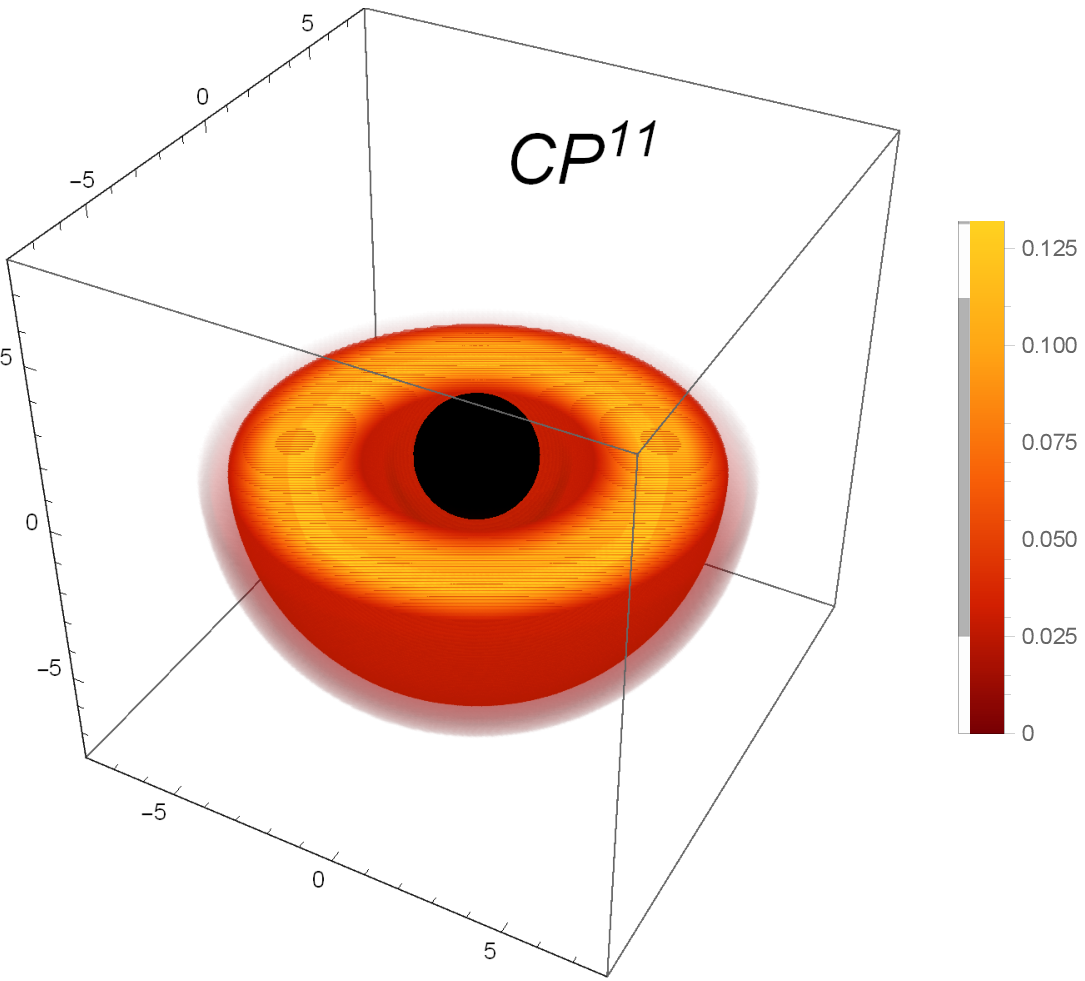} \\
\vspace{5mm}

\includegraphics[width=70mm]{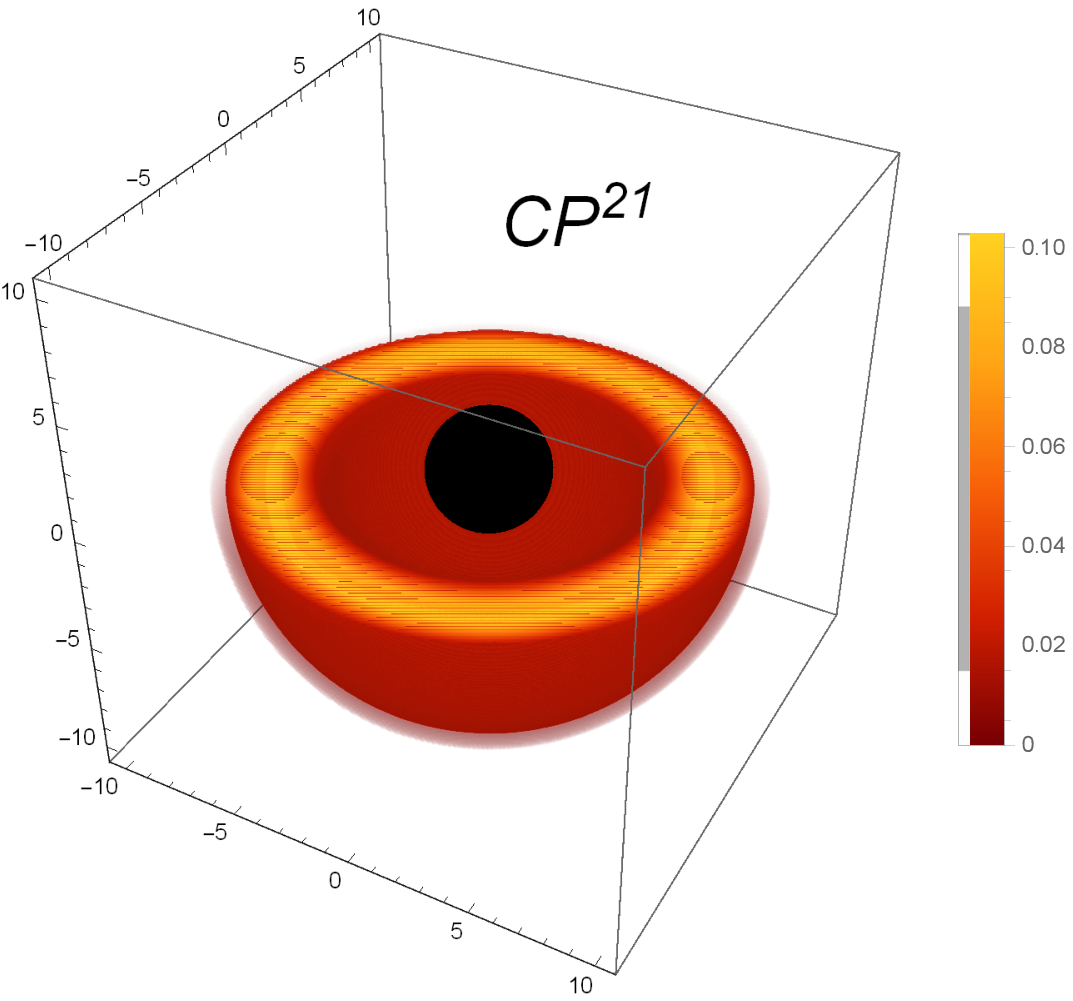}~~~~
\includegraphics[width=70mm]{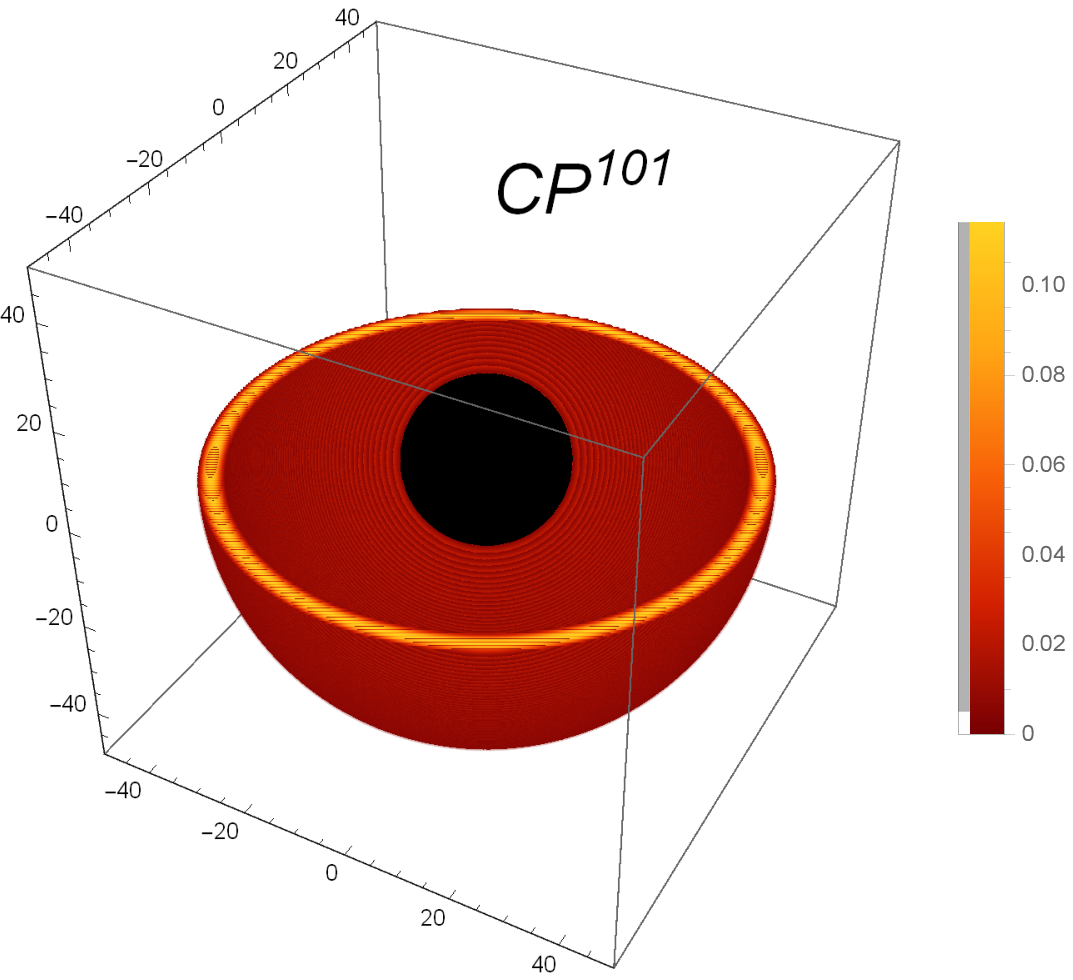} 
\caption{\label{3d} The density of the matter field $|u^\dagger\cdot u|\equiv f(r)^2$ with the black hole (black ball).
The parameters are $\omega=1,0,\alpha=0.01$ for $\mathbb{C}P^5,\mathbb{C}P^{11},\mathbb{C}P^{21}$ 
and $\alpha=0.001$ for $\mathbb{C}P^{101}$.}
  \end{center}
\end{figure*}

	For $n\geqq 2$ we have no nontrivial solutions at the vicinity of the origin $r=0$. 
	It means that the solution is identically zero inside a ball with certain radius. Thus we shall study
	expansion at the sphere with certain finite radius $R_{\rm in}$. Such a solution is known as 
	the $Q$-shell solution because its radial support is restricted to the interval $r\in (R_{\rm in},R_{\rm out})$. 
	The expansions at inner and outer radius of the compacton are very similar. 
	We impose the following boundary conditions at the compacton radius $r=R$
	\begin{align}
	f(R)=0,~~f'(R)=0,~~A(R)=1\,.
	\end{align}
	The functions $f(r)$, $A(r)$ and $m(r)$ are represented by series 
	\begin{align}
	&f(r)=\sum_{k=2}F_k(R-r)^k,~~A(r)=\sum_{k=0}B_k(R-r)^k,\nonumber \\
	&m(r)=\sum_{k=0}M_k(R-r)^k.
	\end{align}
	After determining few first coefficients of expansion we get 
	\begin{align}
	&f(r)=-\frac{R}{16(2M_0-R)}(R-r)^2+\frac{R}{24(2M_0-R)^2}(R-r)^3 \nonumber \\
	&\hspace{5cm}+O((R-r)^4)\,, \nonumber \\
	&A(r)=1-\frac{\alpha R^3}{48(2M_0-R)^2}(R-r)^3+O((R-r)^4)\,, \\
	&m(r)=M_0-\frac{\alpha R^3}{48(2M_0-R)}(R-r)^3+O((R-r)^4)\,. \nonumber 
	\end{align}
	In the region $r>R_{\rm out}$ the metric functions reduce to the 
	Schwarzschild solutions because there is no matter function in this region. Thus we have
	\begin{align}
	m(r)\to M_0,~~A(r)\to 1~~{\rm as}~~r\to R_{\rm out}\,.
	\end{align} 
	It is well-known that certain global black hole observables can be directly deduced 
	from asymptotic expansion of the solutions \citep{Kleihaus:2000kg}. For instance, a mass of black hole
	is defined in terms of $M_0$.

\section{The numerical results}
	\label{result}
In this section we present numerical solutions describing gravitating compact $Q$-balls and $Q$-shells.
We use shooting algorithm to solve differential equations (\ref{einsteinA}), (\ref{einsteinC}), (\ref{matter}) with different $n$ and look at dependence of these solutions on the parameters $\omega$ and $\alpha$ .
According to (\ref{expansion0}) and (\ref{expansion1}) the solutions with $n=0,1$ are regular at the origin. Thus, they are  
$Q$-ball type solutions and they can be designated as boson stars. 
In Fig.\ref{CP1} we plot the solution for the $\mathbb{C}P^1$ case. 
The matter profile function $f(r)$ has non-zero value at the origin and reaches zero 
at the compaction radius. 
We also plot the metric functions $A(r),C(r)$.
With increasing of the coupling constant $\alpha$ the solutions tend to shrink.
The bottom right figure in Fig.\ref{CP1} shows the shooting parameter $f_0$ versus the coupling constant $\alpha$.
We observe the existence of branches of the solutions, i.e., there are two solutions for the unique shooting parameters. 
The solution of the second branch (the dotted line) is stronger localized at the origin.

As discussed in the last section, the solutions can not be nontrivial at the origin for $n\geqq 2$. It leads to existence of $Q$-shells with the matter field localized  in the radial segment $r\in (R_{\rm in},~R_{\rm out})$. 
First we consider the case of self-gravitating solutions, i.e., assume that there is no extra mass at the 
center i.e., $M_0=0$. It means that the space-time in the interior of the $Q$-shell must be regular. 
In Fig.\ref{CPgra} we present examples of typical results for $\mathbb{C}P^5$ and $\mathbb{C}P^{101}$. 
In both cases the metric functions are regular at the origin. The space-time inside empty region of the shell, $r\leqq R_{\rm in}$, 
is Minkowski-like with constant metric functions $A(r)=A_0$ (the dashed lines) and $C(r)=1.0$. 
Our results show that the shell solutions corresponding with higher $n$ become larger and thinner and that the difference 
between self-gravitating $Q$-shells and $Q$-shells in flat space-time is greater for higher $n$. 
We also note that the size of compactons varies significantly with, even small, changes of parameter $\alpha$. 

\begin{figure*}[t]
\begin{center}
\includegraphics[width=120mm]{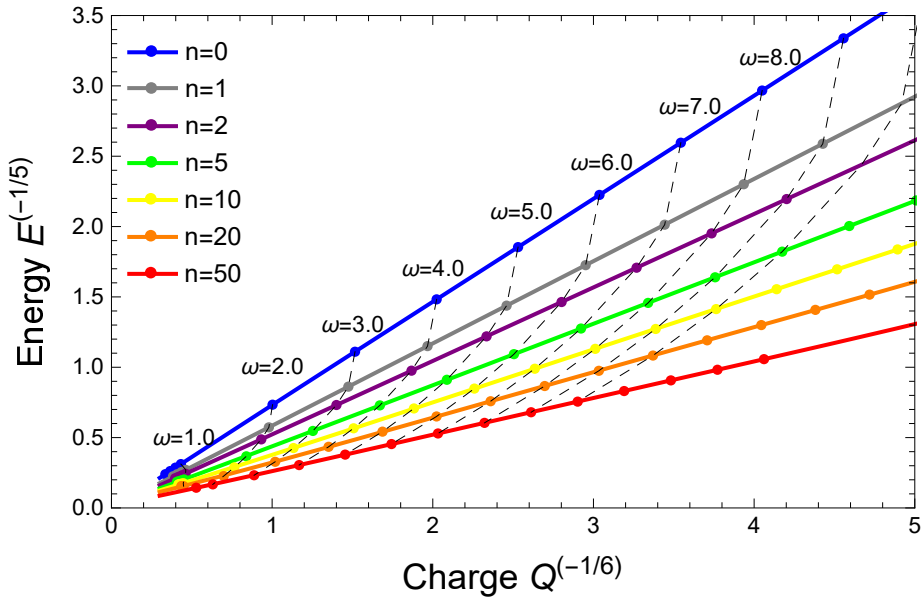}
\caption{\label{EQgrav}The relation between $E^{-1/5}$ and $Q^{-1/6}$ for several gravitating solutions.
The coupling constant takes value $\alpha=0.01$. The dots indicate solutions with different values of $\omega$. }
\end{center}
\end{figure*}

\begin{figure*}[p]
\begin{center}
\includegraphics[width=100mm]{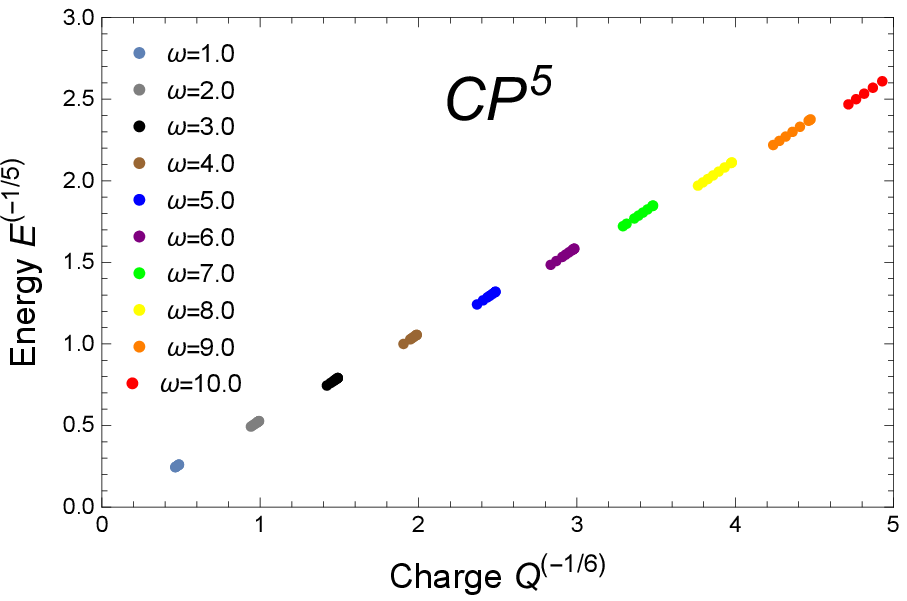}\\
\vspace{0.1cm}

\includegraphics[width=100mm]{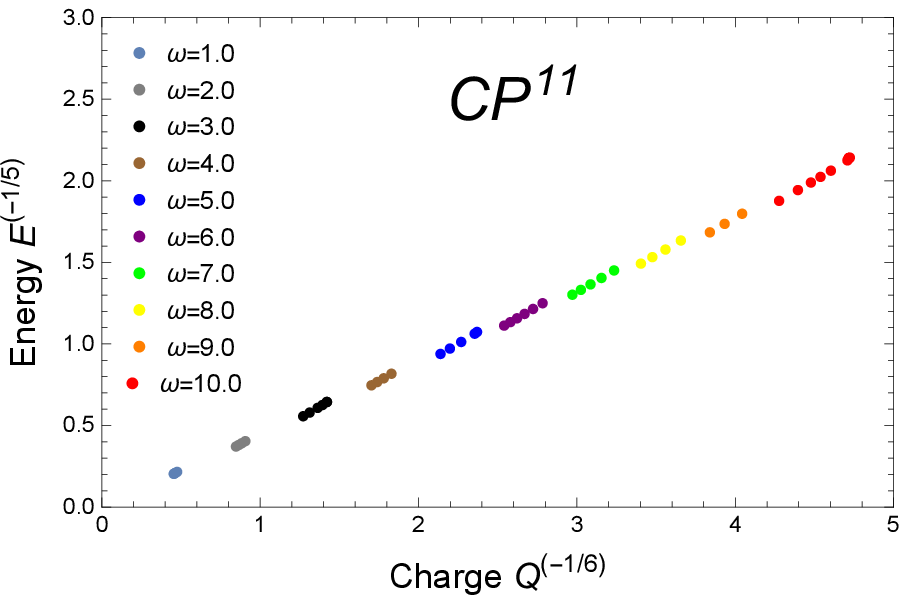}\\
\vspace{0.1cm}

\includegraphics[width=100mm]{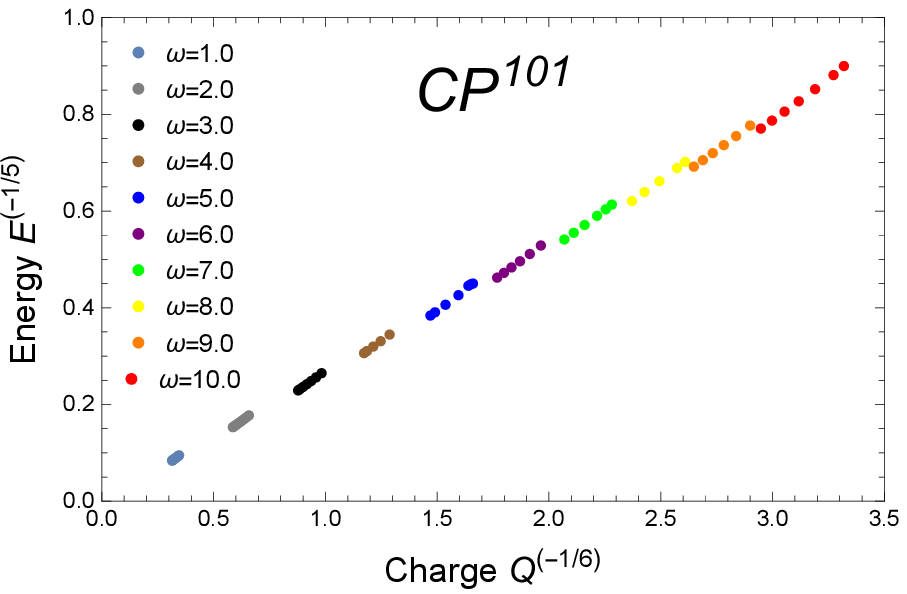}
\caption{\label{EQBH}The relation between $E^{-1/5}$ and $Q^{-1/6}$ for the harbor solutions in the models $\mathbb{C}P^5$,
 $\mathbb{C}P^{11}$ and $\mathbb{C}P^{101}$. The coupling constant takes value $\alpha=0.01$ for $\mathbb{C}P^5$ and
$\mathbb{C}P^{11}$ and value $\alpha=0.001$ for $\mathbb{C}P^{101}$.
The dots with the same color indicate the solutions which differ only by the value of a horizon radius $r_{\rm h}$. 
The largest horizon radius corresponds to the smallest energy in each set of dots. It means that in each set the zero radius 
is represented by the lowest (left) dot and the maximal radius by the highest (right) dot.  
The numerical values of the maximal horizon radius are listed in Table \ref{tab:horizon}.  
}
\end{center}
\end{figure*}

Next we have considered the case of $Q$-shells with a massive body immersed in their center. In 
particular, an interesting possibility is having this body as a  
Schwarzschild-like black hole \citep{Kleihaus:2010ep}. We shall report here on such a case.
We choose the mass $M_0$ such that the event horizon is localized in  a central (empty) part of the shell and 
we solve the equations in the region outside the event horizon. Such solutions are termed harbors. 
The harbor type solutions are obtained in few steps. 
First, we assume that in the region between the event horizon and 
the inner boundary of the shell $r\in [r_{\rm h},R_{\rm in}]$ hold Einstein's vacuum equations as the consequence of 
vanishing of  matter fields in this region. 
A similar approach is used in the region outside the shell $r\in [R_{\rm out},\infty)$, where all matter fields vanish. 
Next, we solve the equations in the region $r\in (R_{\rm in}, R_{\rm out})$ and 
then smoothly connect the metric functions with both vacuum solutions at inner and outer boundary of the shell. 
Note that the values of both   radii of the shell are not {\it a priori} known because they are determined by shooting procedure.   
In Fig.\ref{CPBH} we present typical results for harbor solutions. 
We also present figures which show behavior of the ratio of the radii $r_{\rm h}/R_{\rm in}$ of the event horizon $r_{\rm h}$ and the inner boundary of the shell $R_{\rm in}$.
The ratio $r_{\rm h}/R_{\rm in}=0$ corresponds with the self-gravitating, i.e. $M_0=0$, solutions. 
We have also looked at dependence of the solution on the radius of the event horizon (mass of the black hole). The radii of the shell solution $R_{\rm in}$ and  $R_{\rm out}$ smoothly vary with increasing of $r_{\rm h}$. As $r_{\rm h}$ approaches
$R_{\rm in}$ (the ratio tends to 1) the shell becomes thinner. Also the change of the coupling constant $\alpha$ (see middle figures in Fig.\ref{CPBH}) affects the form of the solution.  We observe that the larger coupling constant is 
 the thinner the shell is. Moreover, we also see that shells again become thinner with increasing the rotational velocity $\omega$.  
A notable difference is visible between the $\mathbb{C}P^{5}$ (the left) and the $\mathbb{C}P^{11}$ (the right).
For the $\mathbb{C}P^{5}$ case the event horizon can be localized very close to the inner border of the shell, however, without touching it exactly. On the other hand, for $\mathbb{C}P^{11}$ the solution collapses before 
the event horizon goes very close to the inner boundary. Also, for larger $n$ the distance between these two radii grows significantly. Therefore we conclude that the black hole can not possesses the hair and the matter field is just a harbor for the black hole. 

Fig.\ref{3d} is quite illustrative and it  shows three-dimensional plots of the harbor solutions where, in concordance with the metric, the event horizon is depicted by black spheres. 
According to the ansatz (\ref{ansatz}) the density of matter fields  possesses spherical symmetry, 
i.e., $|u^\dagger\cdot u|=f(r)^2$. 
The distance between the horizon and the inner boundary grows with increasing of $n$. 
(Note that the reason why the coupling constant was chosen $\alpha=0.001$ for the case $n=50$ was the fact that
we were not able to find solutions above a critical value $\alpha=\alpha_{\rm crit}<0.01$.)

The stability of the $Q$-balls was studied in many literatures, see for instance \citep{Friedberg:1976me,Coleman:1985ki,Lee:1991ax,Sakai:2007ft,Tsumagari:2008bv}. 
An important relation for analysis of the (absolute) stability of classical solutions is the scaling relation of total energy with the Noether charge 
of $Q$-ball or $Q$-shell. This relation was examined analytically in \citep{Klimas:2017eft} where authors study the model (\ref{action}) in the flat space-time. 
In the limit of $\omega\to \infty$ (small amplitude fields) the model reduces to the signum-Gordon model \citep{Arodz:2008jk} which has exact solutions and 
so the relation between the energy and the Noether charge is expected to be $E\sim Q^{5/6}$. 
Unfortunately, for the curved space-time such approximation results in equations which cannot be solved using analytical methods. 
For this reason we study the relation between the energy and the Noether charge numerically and without using any approximations. 
In Fig.\ref{EQgrav} we present  the relation between $E^{-1/5}$ and $Q^{-1/6}$ for self-gravitating solutions in the models with $n=0,1,2,5,10,20,50$.
The dots indicate the solutions with given frequency $\omega$. The energy decreases with increasing of $\omega$. We draw dashed lines which connect the same value of $\omega$ on lines of different $n$. 
According to graphics presented in Fig.\ref{EQgrav} the relation between $E^{-1/5}$ and $Q^{-1/6}$ is linear with very good accuracy so one can conclude that the energy scales with the Noether charge as  $E\sim Q^{5/6}$.  
We have also examined the energy-charge relation for harbor type solutions. 
The results of this analysis are presented in Fig.\ref{EQBH} where we plot $E^{-1/5}$ in function of $Q^{-1/6}$ for $n=2, 5, 50$~
(i.e.,$\mathbb{C}P^5$, $\mathbb{C}P^{11}$, $\mathbb{C}P^{101}$) for several values of $\omega$ and $r_{\rm h}$.
The dots with the same color represent solutions with given $\omega$ and different $r_{\rm h}$. The energy decreases with increasing of the horizon radius. It means that dots of the same color are ordered from $r_{\rm h}=0$ ({\it lower left}) 
to the maximum $r=r_{\rm h,max}$ ({\it upper right}).
The values of $r_{\rm h,max}$ corresponding to this plot are listed in Table \ref{tab:horizon}.
The gradients of the same color are always close to the gradient of the line of $r_{\rm h}=0$ (see Fig.\ref{EQgrav}).
It means that the presence of a black hole is not so apparent for the scaling relation. 
As a result, a linear relation between quantities $E^{-1/5}$ and $Q^{-1/6}$ holds  for both $Q$-ball and $Q$-shell solutions. 

\begin{figure*}[t]
\begin{center}
\includegraphics[width=120mm]{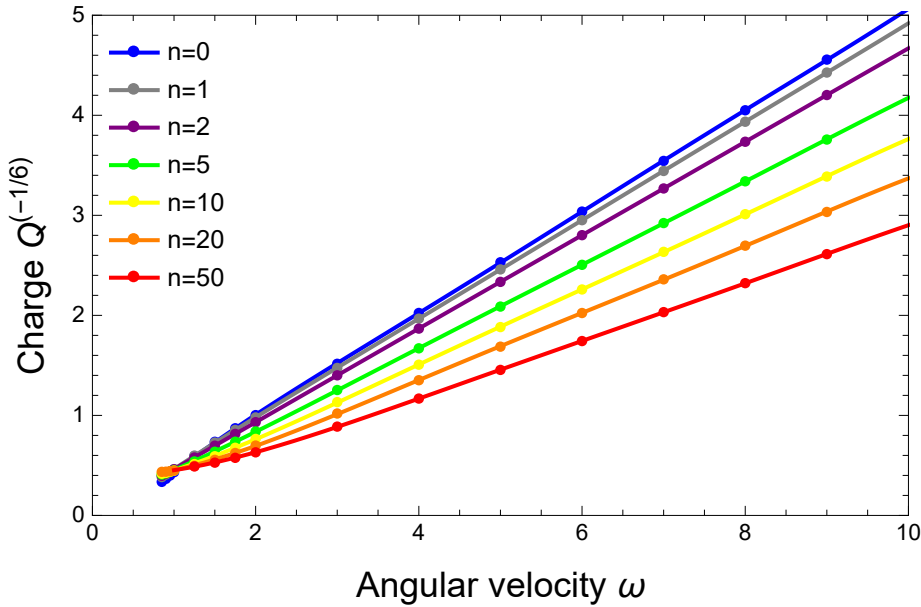}
\caption{\label{wQgrav}The relation between $Q^{-1/6}$ and $\omega$ for several gravitating solutions.
The coupling constant takes value $\alpha=0.01$. The dots indicate solutions with different $\omega$. }
\end{center}
\end{figure*}

In \citep{Klimas:2017eft}, the authors examined the $Q-$$\omega$ relation. They numerically confirmed that it has the form
$Q \sim \omega^{-1/6}$. It thus enjoys a condition of the classical stability, $\frac{1}{Q}\frac{dQ}{d\omega}<0$, proposed in \citep{Friedberg:1976me}.
We have examined this relation for the gravitating solutions. The result is plotted in Fig.\ref{wQgrav}. 
It follows that all solutions with different values of $n$ are stable.

\begin{figure*}[t]
\begin{center}
\includegraphics[width=80mm]{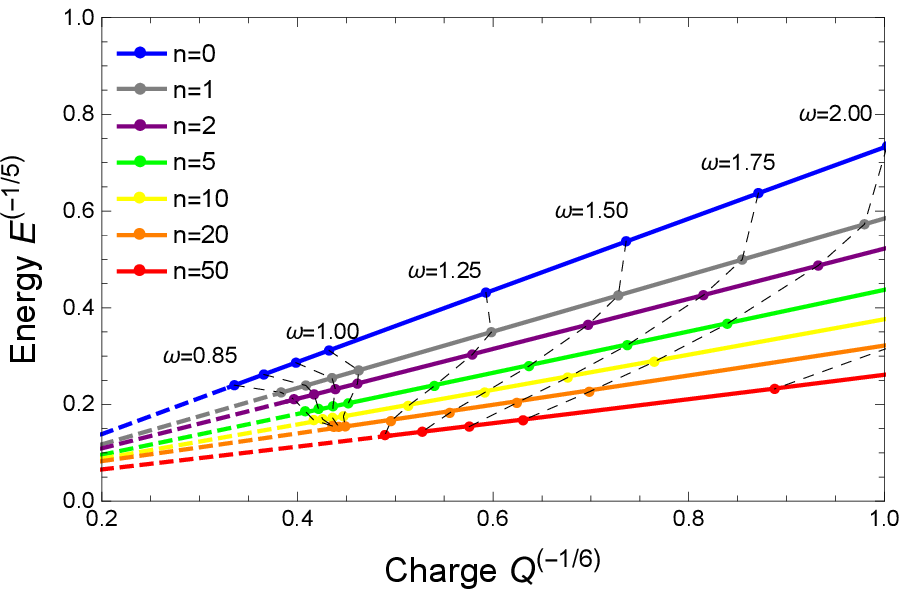}~~~~
\includegraphics[width=80mm]{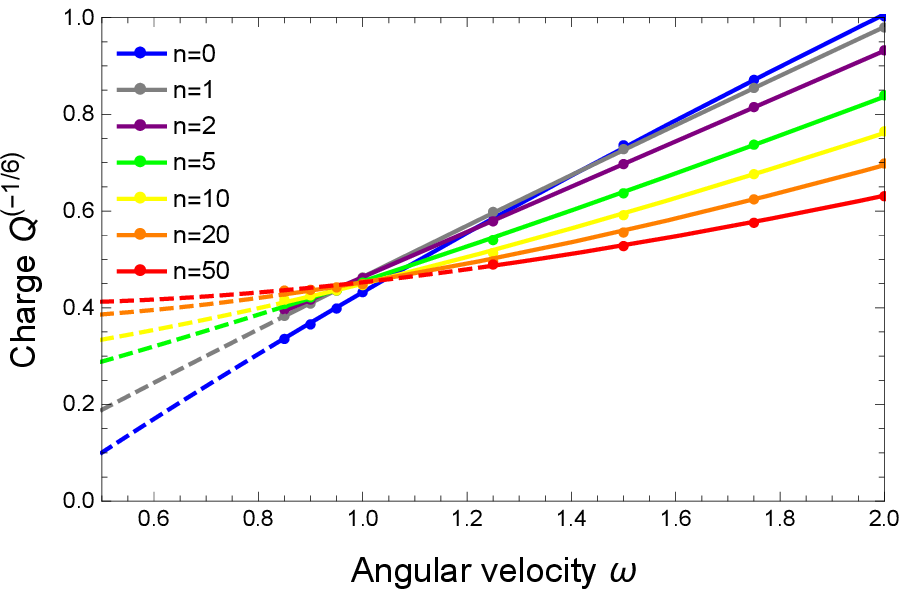}
\caption{\label{small} The relation between $E^{-1/5}$ and $Q^{-1/6}$ for the self-gravitating solutions 
for several $n$ and $\omega$ (left). The relation between $Q^{-1/6}$ and $\omega$ for several $n$ (right).
The dotted lines indicate results of a polynomial extrapolation with sufficient number of terms. }
\end{center}
\end{figure*}

\begin{figure*}[t]
\begin{center}
\includegraphics[width=80mm]{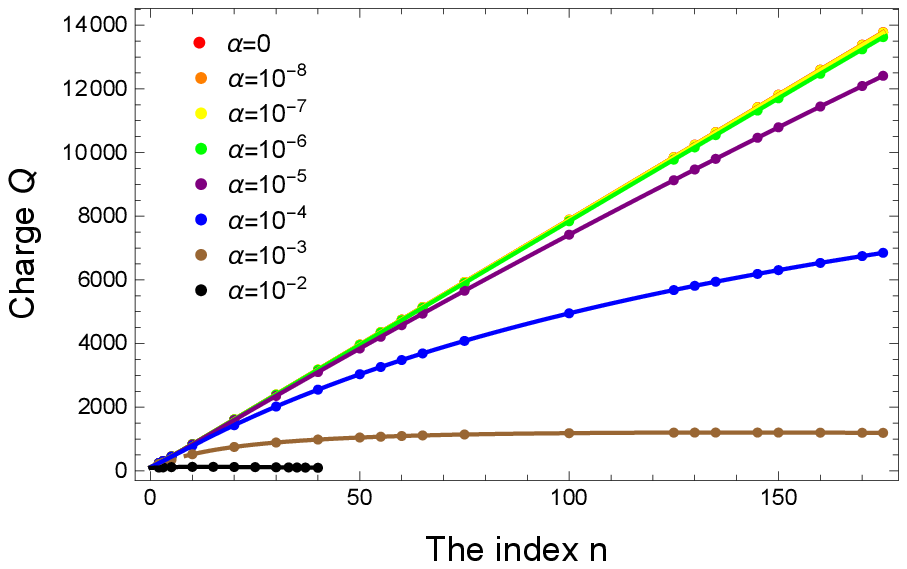}~~~~
\includegraphics[width=76mm]{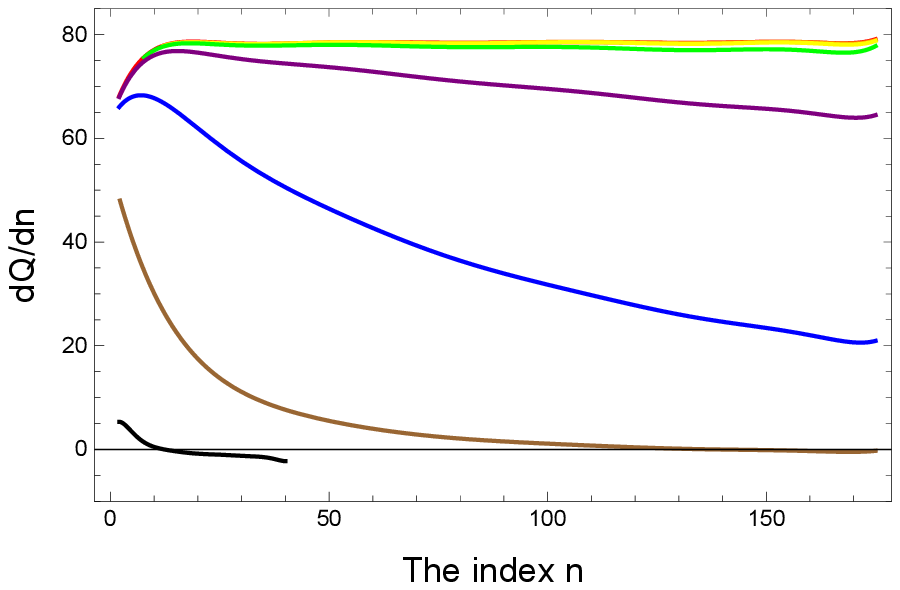}

\caption{\label{nQ}
The Noether charge $Q$ in function of $n$ (left) and the derivative $dQ/dn$ (right)
for $\omega=1.0$. 
}
\end{center}
\end{figure*}

\begin{figure*}[t]
\begin{center}
\includegraphics[width=80mm]{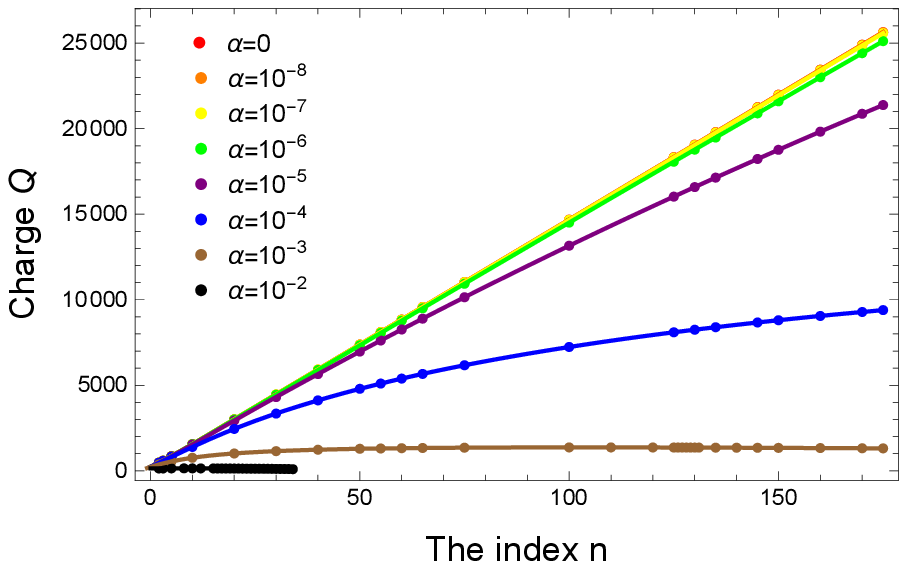}~~~~
\includegraphics[width=76mm]{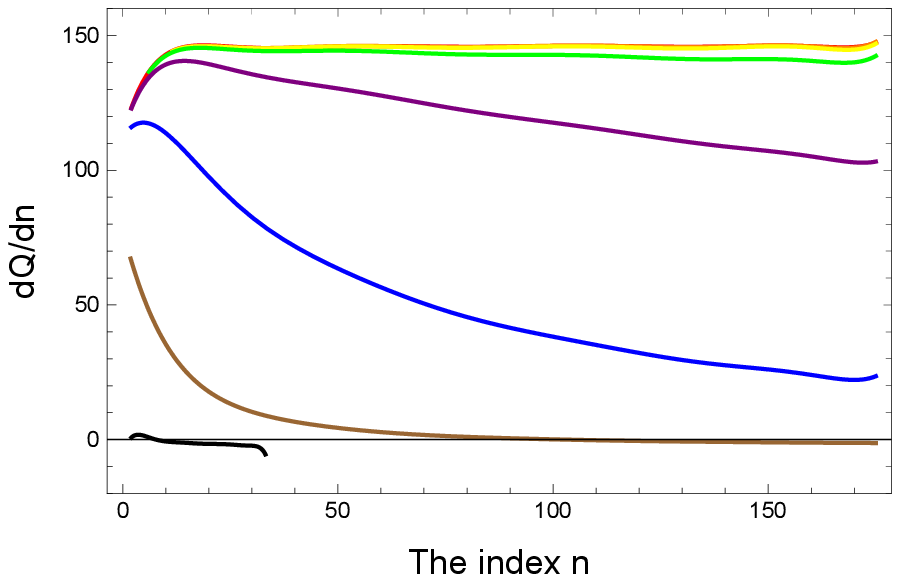}

\caption{\label{nQ2}
Same as the plots of Fig.\ref{nQ} but for $\omega=0.95$. 
}
\end{center}
\end{figure*}

\begin{figure*}[t]
\begin{center}
\includegraphics[width=120mm]{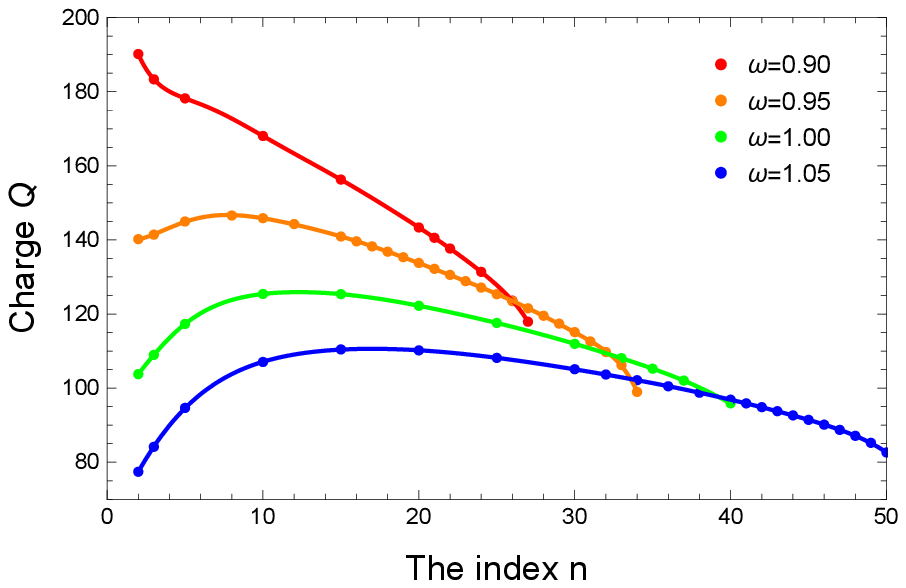}~~~~

\caption{\label{nQcrit}
The Noether charge $Q$ in function of $n$. We plot some cases around $\omega=1.0$ 
with the value $\alpha=10^{-2}$. 
}
\end{center}
\end{figure*}

In our opinion, some further study is required for better understanding of a 
relation between the energy and the Noether charge. The reasons are as follows: 
First, most of the previous studies exhibit some limitations on the frequency $\omega$. Our solutions
were obtained starting with a moderate value $\omega=1$ and then gradually increasing its value. An interesting question is if its value can be taken arbitrary small.   We shall study this question in more detail. 
Second, $Q$-ball solutions reported in the literature have usually no restrictions on maximal value of $|Q|$.
In the gauged $U(1)$ symmetry model, the solutions are unstable for large $|Q|$, which can be overcome 
taking fermions of the scalar condensates with opposite charge, i.e. $-|Q|$, in the interior
and so the charge can reach values around $\sim 1\times 10^8$\citep{Anagnostopoulos:2001dh}. 
Also in the scalar electrodynamics, presented in \citep{Arodz:2008nm}, the authors found $Q$-balls with scaling relation $E\sim |Q|^{7/6}$ for large $Q$. This result may indicate instability of solutions for large $Q$ and their tendency to evaporate.  
In \citep{Klimas:2017eft}, the authors do not report on possible existence of upper limit for $Q$. 
(Instead, they discuss the scaling relation based on the signum-Gordon limit of the model $\omega\to \infty$.) 
Thus there is certain variety of contexts in which the relations between $Q$,  $\omega$ and $E$ are presented.
These analyses were conducted in the flat space-time, i.e., the effects caused by gravity has always been ignored. 
Certainly it is worth to investigate deeper such relations for the systems where the gravity is taken into account. 
One of our motivation for such a study is checking if  there can exist an upper limit for the charge $|Q|$ of gravitating systems. 
In the next section, we shall report on this issue in detail.

\begin{table}[t]
\caption{The maximal value of the horizon radius for the values of $\omega$.}
  \begin{tabular}{lrrr}
\hline\hline
$\omega$~~~~~~~~& \multicolumn{3}{c}{$r_{\rm h,max}$}  \\
        &~~$\mathbb{C}P^5$ &~~$\mathbb{C}P^{11}$&~~$\mathbb{C}P^{101}$ \\
\hline
    1.0 & ~~~~0.5385 & ~~~~1.6814 & ~~~~14.752\\
    2.0 & ~~~~0.2851 & ~~~~1.0647 & ~~~~10.256\\
    3.0 & ~~~~0.1902 & ~~~~0.7188 & ~~~~6.959 \\
    4.0 & ~~~~0.1427 & ~~~~0.5402 & ~~~~5.235\\
    5.0 & ~~~~0.1142 & ~~~~0.4324 & ~~~~4.192\\
    6.0 & ~~~~0.09513 & ~~~~0.3604 & ~~~~3.494\\
    7.0 & ~~~~0.08155 & ~~~~0.3090 & ~~~~2.995\\
    8.0 & ~~~~0.07137 & ~~~~0.2703 & ~~~~2.621\\
    9.0 & ~~~~0.06344 & ~~~~0.2403 & ~~~~2.330\\
   10.0 & ~~~~0.05705 & ~~~~0.2163 & ~~~~2.097 \\
\hline\hline
  \end{tabular}
\label{tab:horizon}
\end{table}

	\section{Further analysis}
	\label{limit}
	
According to results reported in the literature the frequency $\omega$ is restricted to a certain finite interval. Compact $Q$-balls/shells are rather exceptional in this context because 
their existence do not require such restrictions. In fact, for a $V$-shaped potential there is no upper bound for  $\omega$ and the lower bound of $\omega$ is zero (for the signum-Gordon model) or given by certain positive constant (the case of model \eqref{lag0}). In both cases admissible frequencies $\omega$  belong to infinite intervals. For this reason we shall look in more detail at the region of Fig.\ref{EQgrav} corresponding with $\omega\le 2.0$. 
The results of our analysis are shown in Fig.\ref{small}. In the case scaling relation between $E^{-1/5}$ and $Q^{-1/6}$ the linearity holds strictly even without assuming that $\omega\to \infty$. A more interesting relation is that between
$Q^{-1/6}$ and $\omega$. For small $n$, the relation is linear, however, as 
$n$ grows, its exhibits distortion from the linearity and simultaneously the lower bound of 
$\omega$ is gradually shifted in direction of higher values of $\omega$. The derivative $\frac{dQ^{-1/6}}{d\omega}$ decreases with increasing of $n$. In particular, for $n=50$ the derivative approaches zero for small $\omega$. 
This means that the condition for classical stability is not fulfilled any longer. It indicates the existence of a lower bound of $\omega$ for gravitating solutions.

Now we would like present some arguments which would shed light on the role of number of fields $N:=2n+1$. 
In Fig.\ref{nQ} and Fig.\ref{nQ2}, we plot the Noether charge $Q$ in dependence on the number $n$ for 
several coupling constants $\alpha$ and two values of the frequency $\omega=1.0$ and $\omega=0.95$. 
For better transparency we plot also the interpolating function which depends on a continuous variable  (also called $n$).
In the case of flat space-time, $\alpha=0$, the relation is almost linear, which
means that the model with given $N$ has a solution with corresponding particle number $Q$. 
Therefore, both $Q$ and $N$ indicate the particle number of the constituents. 
In right figures we present plots of the derivative $\frac{dQ}{dn}$ for the interpolating functions.   
Notable property of solutions in the curved space-time is that the correspondence between $N$ and $Q$
is broken. 
It means that the increment of $n$ by one causes in generality a moderate change of the charge $Q$. 
However, at critical value of $\alpha$ the charge $Q$ has maximum (which is seen from vanishing of $\frac{dQ}{dn}$ at this point), 
and for larger $n$ the charge $Q$ decreases with increasing of $n$.
The peak moves to the left and became higher with decreasing of $\omega$.  When $\omega$ grows it moves to the right and became lower. 
In Fig.\ref{nQcrit}, we draw for some cases of $\omega$ for $\alpha=10^{-2}$ and one can easily confirm the behavior. 
We conclude that above the critical value of the coupling constant $\alpha$ 
there is a maximal value of the charge $Q_{\rm max}$. Although this value became larger for smaller values of $\omega$ the system eventually tends to instability as discussed above. We conclude that the 
gravitating solutions above the critical coupling have also an upper bound for $|Q|$.

It may give us a hint how a boson star with certain number of particles forms its shape. 
The systems with large $n$ can be achieved via a weak coupling constant $\alpha$ and also a small frequency $\omega$.

	\section{Summary}
	\label{summary}

In this paper, we have considered the family of $\mathbb{C}P^N$ nonlinear sigma models 
coupled with gravity which possess  solutions with compact support.  Such solutions, in form of compact $Q$-balls and  $Q$-shells, were obtained solving numerically the system of coupled equations. The self-gravitating solutions of this kind are interpreted as boson stars.
We have also considered the case of spherical $Q$-shells with the Schwarzschild-like black hole immersed in their center. The space-time in the region where hold the Einstein's vacuum equations (interior end exterior of the shell) is described by the Schwarzschild metric. 
Although such solutions are possible candidates which would contradict the no-hair conjecture, 
we have shown that in its strict sense, this is not the case of our solution.
Notable difference from solutions in the flat space is the existence of 
upper bound for $|Q|$ for coupling constants $\alpha$ bigger than certain critical value.  We have checked that the value of this bound drastically decreases with increasing of gravitational coupling constant. 
For sufficiently small coupling constant, the maximal value of the $|Q|$ is reached for large $n$, 
It might be a possible explanation why the boson stars have definite size and mass. 

We restricted our analysis to the case of asymptotically flat Schwarzschild space-time. We managed to obtain the 
stable shell solutions for $n\geqq 2$ even though the matter fields are not coupled with electromagnetism.
The the model (\ref{lagdimless}) possesses symmetry $SU(2)\otimes U(1)$.  
In current research we just focused on the subgroup $U(1)^N$ of the full symmetry group. 
Further, the ansatz (\ref{ansatz}) reduces the symmetry to $U(1)$ and the corresponding Noether charge 
plays central role in stability of our solutions. 

This paper constitutes an initial step for construction of gravitating $Q$-balls (-shells) with non-Abelian symmetry $SU(2)\otimes U(1)$. 
The following problems require solution: 
\begin{itemize}
\item 
Extensions of our model onto the model with finite cosmological constants, i.e. onto the de Sitter or anti-de Sitter space-time 
background. Such extension is almost straightforward and it will be reported in the subsequent paper. 

\item
A coupling with electromagnetism which will allow us to treat the charged boson stars and field configurations  
with the Reissner-Nordstr\"om-type metric. The analysis of these models is in progress.. 

\item 
A study of non-Abelian Noether charges. There are know $Q$-balls in scalar field models with non-Abelian symmetry \citep{Safian:1987pr,Safian:1988cz}. 
The boson stars with a non-Abelian SU(2) symmetry are discussed in \citep{Brihaye:2004nd,Giacomini:2017xno}. 
The existence of $Q$-balls with the symmetry $SU(2)\otimes U(1)$ is highly probable. 
It would be an interesting possibility because it would two types of Noether charges which have crucial role for the stabilization of 
nontopological solitons. 

\end{itemize}

We shall report on these issues in near future.

\vskip 0.5cm\noindent
\begin{center}
{\bf Acknowledgment}
\end{center}

	The authors would like to thank L. A. Ferreira for discussion and many helpful advices. 
	N.S. and S.Y. thank Yuki Amari, Satoshi Horihata, Takanori Kodama, Atsushi Nakamula, Ryu Sasaki, 
	Kouichi Toda for many helpful discussions.
	We are  grateful to Betti Hartmann for her kind hospitality at the 
	``Workshop on Solitons: Integrability, Duality and Applications''.
	N.S. also appreciates  useful discussions with Yakov Shnir and the kind 
	hospitality of his institute where part of this work was done.
	S.Y. thanks the Yukawa Institute for Theoretical Physics at Kyoto University. 
	Discussions during the YITP workshop YITP-T-18-04 on "New Frontiers in String 
	Theory 2018" were useful to complete this work.	
	N.S. was supported in part by JSPS KAKENHI Grant Number JP 16K01026.

	\bibliographystyle{apsrev4-1.bst}
	\bibliography{cpngqshell}

\end{document}